\documentclass{revtex4}
\usepackage{eurosym}
\usepackage[utf8]{inputenc}
\usepackage{amsmath}
\usepackage{amsfonts}
\usepackage{t4phonet}
\usepackage{amssymb}
\usepackage{setspace}
\usepackage{tipa}
\usepackage{epsfig}
\usepackage{graphicx}
\usepackage{float}
\usepackage{subfigure}

\begin{document}

\title{Singular and regular vortices on top of a background pulled to the
center}
\author{Zhaopin Chen$^{1}$ and Boris A. Malomed$^{1,2}$}
\affiliation{$^{1}$Department of Physical Electronics, School of Electrical Engineering,
Faculty of Engineering, and Center for Light-Matter Interaction, Tel Aviv
University, P.O.B. 39040, Tel Aviv, Israel\\
$^2$Instituto de Alta Investigaci\'{o}n, Universidad de Tarapac\'{a},
Casilla 7D, Arica, Chile}

\begin{abstract}
A recent analysis has revealed singular but physically relevant localized 2D
vortex states with density $\sim r^{-4/3}$ at $r\rightarrow 0$ and a
convergent total norm, which are maintained by the interplay of the
potential of the attraction to the center, $\sim -r^{-2}$, and a
self-repulsive quartic nonlinearity, produced by the Lee-Huang-Yang
correction to the mean-field dynamics of Bose-Einstein condensates. In
optics, a similar setting, with the density singularity $\sim r^{-1}$, is
realized with the help of quintic self-defocusing. Here we present
physically relevant \textit{antidark} singular-vortex states in these
systems, existing on top of a flat background. Numerical solutions for them
are very accurately approximated by the Thomas-Fermi wave function. Their
stability exactly obeys an analytical criterion derived from analysis of
small perturbations. The singular-vortex states exist as well in the case
when the effective potential is weakly repulsive. It is demonstrated that
the singular vortices can be excited by the input in the form of the
ordinary nonsingular vortices, hence the singular modes can be created in
the experiment. We also consider regular (dark) vortices maintained by the
flat background, under the action of the repulsive central potential $\sim
+r^{-2} $. The dark modes with vorticities $l=0$ and $1~$are completely
stable. In the case when the central potential is attractive, but the
effective one, which includes the centrifugal term, is repulsive, and, in
addition, a weak trapping potential $\sim r^{2}$ is applied, dark vortices
with $l=1$ feature an intricate pattern of alternating stability and
instability regions. Under the action of the instability, states with $l=1$
travel along tangled trajectories, which stay in a finite area defined by
the trap. The analysis is also reported for dark vortices with $l=2$, which
feature a complex structure of alternating intervals of stability and
instability against splitting. Lastly, simple but novel \textit{flat vortices%
} are found at the border between the anidark and dark ones.
\end{abstract}

\maketitle

\section{Introduction}

In typical cases, vortex states exist, in self-defocusing optical media \cite%
{Swartz}-\cite{opt-review3} and Bose-Einstein condensates (BECs, i.e., in
optics of coherent matter waves) \cite{BEC-review1}-\cite{BEC-review3}, as
two-dimensional (2D) dark solitons, supported by a modulationally stable
background. In terms of polar coordinates $\left( r,\theta \right) $, the
wave function of a vortex with integer winding number (topological charge) $%
l\geq 0$ has the standard asymptotic form at $r\rightarrow 0$:
\begin{equation}
\psi \approx \mathrm{const}\cdot r^{l}e^{il\theta }.  \label{r^l}
\end{equation}%
Unlike this classical situation, physically relevant \textit{singular
vortices}, as well as zero-vorticity states, were predicted in 2D models
which combine the attractive potential
\begin{equation}
U(r)=-U_{0}/\left( 2r^{2}\right) ,  \label{U}
\end{equation}%
with $U_{0}>0$, and a self-repulsion term in the underlying Gross-Pitaevskii
(GP) \cite{Pit} or nonlinear Schr\"{o}dinger (NLS)\ equation, which must be
stronger than cubic, i.e.,
\begin{equation}
\mathrm{NLS~term}=|\psi |^{\alpha }\psi  \label{alpha}
\end{equation}
with $\alpha >2$ \cite{HS1,we,small-review}. As demonstrated in \cite{HS1},
potential (\ref{U}) in the GP equation for a molecular or atomic gas may be
provided by attraction of a small molecule with a permanent electric dipole
moment to a central charge, or attraction of a magnetically polarizable atom
to electric current (such as an electron beam) piercing the plane of the 2D
system (a similar but weaker singular attractive potential, $\sim -1/r$, was
considered under the name of a funnel potential in works \cite{funnel} and
\cite{funnel2}, but it cannot create singular vortices). In optics,
effective potential (\ref{U}) may be realized by dint of resonant dopants,
with a spatially modulated resonance detuning \cite{Wesley}. The modulation
can be imposed by a non-uniform external magnetic or electric field, via the
Zeeman or Stark - Lo Surdo effect, respectively \cite{LL}.

In such a setting, there are solutions, singular at $r\rightarrow 0$, with
asymptotic form (\ref{r^l}) replaced by
\begin{equation}
\psi \approx \left[ \frac{1}{2}\left( U_{l}+\frac{4}{\alpha ^{2}}\right) %
\right] ^{1/\alpha }r^{-2/\alpha }e^{-i\mu t+il\theta },  \label{singular}
\end{equation}%
where $\mu $ is the chemical potential, and an effective strength of the
pull to the center including a centrifugal term:%
\begin{equation}
U_{l}\equiv U_{0}-l^{2}.  \label{Ul}
\end{equation}%
Solution (\ref{singular}) exists under the condition of $U_{l}+4/\alpha
^{2}>0$, see further details below. Although this solution is singular, it
is a physically acceptable one, as the respective norm,
\begin{equation}
N=2\pi \int_{0}^{\infty }\left\vert \psi (r)\right\vert ^{2}rdr,  \label{N}
\end{equation}
converges at $r\rightarrow 0$ under the condition of $\alpha >2$, i.e., the
self-repulsion must be stronger than cubic. In work \cite{HS1}, the
existence of such normalizable singular vortex modes was demonstrated for $%
\alpha =4$ in Eq. (\ref{alpha}), which corresponds to the quintic
self-defocusing, well known in diverse forms in optics \cite%
{Michinel,Leblond,Boyd,Jana,Cid}, or the effect of three-body collisions in
BEC \cite{Abdullaev}. Then, a similar result was reported in work \cite{we}
for the GP equation including the beyond-mean-field Lee-Huang-Yang (LHY)
correction, produced by an effect of quantum fluctuations in binary BEC. The
latter term amounts to the quartic nonlinearity, with $\alpha =3$ in Eq. (%
\ref{alpha}) \cite{Petrov1,Petrov2,2D1,2D2}. This nonlinearity corresponds
to experiments producing stable self-trapped \textit{quantum droplets }\cite%
{Leticia1}-\cite{hetero}. In additional to the systematic numerical
investigation, exact analytical results which determine a stability boundary
for the singular vortices, and approximate results for the shape of the
vortices, were also reported in \cite{we}.

It is relevant to mention that the three-dimensional (3D) GP-like equation
with the same attractive potential (\ref{U}) and the self-repulsive term (%
\ref{alpha}) has fundamental (zero-vorticity) isotropic solutions with the
singular asymptotic form \cite{HS1}%
\begin{equation}
\psi _{\mathrm{3D}}\approx \left[ \frac{1}{2}\left( U_{0}+\frac{2}{\alpha }%
\left( \frac{2}{\alpha }-1\right) \right) \right] ^{1/\alpha }r^{-2/\alpha
}e^{-i\mu t}.  \label{3D}
\end{equation}%
cf. Eq. (\ref{singular}). The convergence of the respective 3D norm,
\begin{equation}
N_{\mathrm{3D}}=4\pi \int_{0}^{\infty }\left\vert \psi (r)\right\vert
^{2}r^{2}dr,
\end{equation}
at $r\rightarrow 0$ is secured by condition $\alpha >4/3$. The nonlinearity
with $\alpha =4/3$ corresponds to the effective self-repulsion in the
quantum Fermi gas, according to the density-functional approximation \cite%
{Fermi1}-\cite{Fermi3}, but it still produces a weak logarithmic divergence
of the 3D norm. Obviously, the physically relevant cubic, quartic, and
quintic nonlinearities all satisfy the convergence condition, while there is
no physically relevant example of the nonlinearity with $4/3<\alpha <2$.

A noteworthy feature of the numerical and analytical results presented in
\cite{we} for the 2D setting is the fact that the singular vortices exist
not only in the case of $U_{l}>0$ (see Eq. (\ref{Ul})), when they are
maintained by the pull to the center, but also in the interval of
\begin{equation}
-4/9<U_{l}<0  \label{4/9}
\end{equation}%
(in terms of the notation adopted below in Eq. (\ref{basic})), where the
effective potential is slightly repulsive. In particular, for $l\geq 1$ the
underlying potential (\ref{U}), corresponding to interval (\ref{4/9}), is
still attractive, while it is repulsive, indeed, for the zero-vorticity
modes, with $l=0$. This counter-intuitive finding can be explained by the
fact that the NLS equation with the self-repulsive nonlinearity may give
rise to bright singular-solitons solutions \cite{Veron,HS2}. It is also
relevant to mention, in this connection, that, in the 3D case with the cubic
self-repulsion ($\alpha =2$), the physically acceptable singular state (\ref%
{3D}) exists precisely at $U_{0}>0$, while the quartic or quintic terms ($%
\alpha =3$ or $4$ in Eq. (\ref{alpha})) provide for the convergence of the
norm if the attraction strength exceeds a finite threshold, \textit{viz}., $%
U_{0}>2/9$ or $1/4$, respectively.

It is relevant to mention that the concept of vortices with singular cores
is also known in theoretical studies of the 2D dynamics in spinor
(three-component) BEC \cite{Japan,UK}. However, the meaning of such modes is
different in that context, as their densities remain finite, while the
singularity implies splitting of vorticity axes of the three components.

In works \cite{HS1} and \cite{we}, the singular zero-vorticity and vortex
states in the 2D\ system combining potential (\ref{U}) and the quintic or
LHY nonlinearity were found, respectively, as ones localized at $%
r\rightarrow \infty $, i.e., with a negative chemical potential, $\mu <0$. A
natural possibility, which is the subject of the present work, is to
construct physically relevant 2D singular states, especially vortical ones,
featuring the asymptotic form (\ref{singular}), on top of a modulationally
stable background with a finite density at $r\rightarrow \infty $, which
corresponds to $\mu >0$. In that sense, these states may be called \textit{%
antidark} ones \cite{antidark1}-\cite{antidark3}. Parallel to the
consideration of them, we also produce usual (regular) vortices of the dark
type, i.e., solutions subject to the asymptotic form (\ref{r^l}) at $%
r\rightarrow 0$, and quite simple but novel solutions in the form of \textit{%
flat vortices}, which exist at the border between the antidark and dark
vortex modes.

The subsequent presentation is organized as follows. The model is formulated
in Section II, where we also present some analytical results\ -- in
particular, those obtained by means of the Thomas-Fermi (TF) approximation.
Numerical results for the existence and stability of singular and regular
(antidark and dark, respectively) states are reported in Section III. An
essential novel numerical result concerns the excitation of the singular
vortex from an input, which may only be a regular (nonsingular) optical
vortex in free space. This issue, which is crucially important for
predicting the possibility to create the singular vortex in the experiment,
was not addressed in previous works. Direct simulations reported in Section
III clearly demonstrate that stable singular vortices are readily excited by
nonsingular inputs carrying the vorticity. The paper is concluded by Section
V.

\section{Basic equations and analytical approximations}

\subsection{The nonlinear-Schr\"{o}dinger/Gross-Pitaevskii equation}

The underlying 2D NLS/GP\ equation for the mean-field wave function, $\psi $%
, including the above-mentioned ingredients, i.e., potential (\ref{U}) and
the nonlinear term (\ref{alpha}), was introduced in works \cite{HS1} and
\cite{we}. Here we write the scaled equation in the polar coordinates:%
\begin{eqnarray}
i\frac{\partial \psi }{\partial t} &=&-\frac{1}{2}\left( \frac{\partial
^{2}\psi }{\partial r^{2}}+\frac{1}{r}\frac{\partial \psi }{\partial r}+%
\frac{1}{r^{2}}\frac{\partial ^{2}\psi }{\partial \theta ^{2}}\right)  \notag
\\
&&-\frac{U_{0}}{2r^{2}}\psi +|\psi |^{\alpha }\psi +\frac{k}{2}r^{2}\psi .
\label{basic}
\end{eqnarray}%
We focus on the most essential case when Eq. (\ref{basic}) does not include
a cubic term. In particular, the optical nonlinearity of the colloidal
suspension of metallic nanoparticles can be accurately adjusted so as to
eliminate the cubic part, keeping only the quintic one, with $\alpha =4$ in
Eq. (\ref{basic}). In BEC, the cubic mean-field intra-component
self-repulsion and inter-component attraction in the binary condensate may
be brought in full balance, the quartic term ($\alpha =3$) being the single
nonlinear one in the system (the \textit{LHY liquid} \cite{LHY-only}).

As concerns the BEC\ realization, the limit of extremely tight confinement
in the transverse direction, which provides the reduction of the underlying
3D GP equation to the effective 2D form, corresponds to the case when the
confinement size, $a_{\perp }$, is much smaller than the BEC\ healing
length, $\xi $. In this limit, the nonlinearity in the LHY-corrected 2D GP
equation is different from that given by Eq. (\ref{alpha}). Instead, it is\
written as $\ln \left( |\psi |^{2}\right) \cdot |\psi |^{2}\psi $ \cite%
{Petrov2,2D1,2D2}. On the other hand, experiments are usually conducted
under the opposite condition, $\xi \ll a_{\perp }$. In this case, one can
still use the quartic nonlinearity in Eq. (\ref{basic}), assuming that a
characteristic lateral size of patterns produced by this equation is much
larger than $a_{\perp }$, which is a realistic condition \cite{we}.

Equation (\ref{basic}) includes the trapping harmonic-oscillator (HO)
potential with strength $k\geq 0$. Aiming to consider modes supported by the
flat background, the trap may be dropped. In terms of the experimental
realization, which always includes the HO trap \cite{Pit} in BEC, or a
cladding which confines the waveguide in optics, that may be approximated by
the last term in Eq. (\ref{basic}) \cite{Agrawal}, setting $k=0$ means that
the characteristic size of the mode's core, which is either the singular
peak of the anti-dark vortex, or the \textquotedblleft hole" of the regular
dark one, is localized in an area of a size $\ll k^{-1/2}$. Nevertheless,
the HO potential plays an essential role in the analysis of stability of
dark vortices \cite{HanPu}, which is also shown below.

To address, as said above, both singular (antidark) and regular (dark)
modes, we here consider positive and negative values of $U_{0}$ in Eq. (\ref%
{basic}). In optics the sign of $U_{0}$ is determined by the sign of the
detuning between the carrier electromagnetic wave and resonant dopants. In
terms of the above-mentioned realization of the setup in BEC, based on the
magnetic polarizability of atoms, $U_{0}<0$ corresponds to diamagnetic
susceptibility \cite{diamagnetic1,diamagnetic2}.

Stationary solutions with chemical potential $\mu >0$ and integer vorticity $%
l\geq 0$ (a.k.a. the photonic angular momentum, in terms of the
corresponding optics models \cite{opt-review3}) are looked for as%
\begin{equation}
\psi \left( r,t\right) =u(r)e^{-i\mu t+il\theta },  \label{psi-u}
\end{equation}%
where real radial function $u(r)>0$ satisfies equation%
\begin{eqnarray}
\mu u &=&-\frac{1}{2}\left( \frac{d^{2}u}{dr^{2}}+\frac{1}{r}\frac{du}{dr}+%
\frac{U_{l}}{r^{2}}u\right)  \notag \\
&&+u^{\alpha +1}+\frac{k}{2}r^{2}u.  \label{u}
\end{eqnarray}%
Obviously, for $\mu >0$ Eq. (\ref{u}) with $k=0$ admits the presence of the
modulationally stable flat background,%
\begin{equation}
u^{2}(r\rightarrow \infty )=\mu ^{2/\alpha }.  \label{u0}
\end{equation}%
The respective asymptotic form of the solution to Eq. (\ref{u}) at $%
r\rightarrow \infty $ is%
\begin{equation}
u=\mu ^{1/\alpha }+\frac{U_{l}}{2\alpha \mu ^{1-1/\alpha }}r^{-2}+O\left(
r^{-4}\right) .  \label{large r}
\end{equation}

For $U_{l}+4/9>0$, the asymptotic form of the singular solution to Eq. (\ref%
{u}) at $r\rightarrow 0$, which extends the above expression (\ref{singular}%
), is
\begin{equation}
u=\left[ \frac{1}{2}\left( U_{l}+\frac{4}{\alpha ^{2}}\right) \right]
^{1/\alpha }r^{-2/\alpha }+\frac{2^{1-1/\alpha }}{3}\mu \frac{\left(
U_{l}+4/\alpha ^{2}\right) ^{1/\alpha }}{\alpha U_{l}+4\left( 3/\alpha
-1\right) }r^{2-2/\alpha }+O\left( r^{10/3}\right) .  \label{sigma=0}
\end{equation}%
%
In the special case of the quintic nonlinearity ($\alpha =4$) and $U_{l}=1/4$%
, the correction term in Eq. (\ref{sigma=0}) diverges, being replaced by one
$\sim \sqrt{\mu r}$. Similarly, the correction term diverges in the case of
the quartic nonlinearity ($\alpha =3$) and $U_{l}=0$, being replaced by a
term $\sim \sqrt{\mu }r^{1/3}$. It is relevant to mention that the
linearized version of Eq. (\ref{basic}) produces no counterpart of this
asymptotic solution, hence it does not bifurcate from any solution of a
linear equation.

Note that, in the special case of $U_{l}=0$, i.e., $U_{0}=l^{2}$ (see Eq. (%
\ref{Ul})), Eq. (\ref{u}) with $\mu >0$ and $k=0$ admits a simple but novel
solution, which features a flat density profile but, nevertheless, carries
the vorticity. This solution can be written in terms of the polar
coordinates, as well as in the Cartesian coordinates, $\left( x,y\right) $:

\begin{equation}
\psi _{\mathrm{flat}}=\mu ^{1/\alpha }e^{-i\mu t+il\theta }\equiv \mu
^{1/\alpha }e^{-i\mu t}\left( \frac{x+iy}{x-iy}\right) ^{l/2}.  \label{flat}
\end{equation}%
The flat-vortex state is an intermediate one between the singular (antidark)
and regular (dark) ones.

It is relevant to mention that, for $U_{l}<0$, the linearized version of Eq.
(\ref{u}) admits an \emph{exact} ground-state (GS)\ solution,%
\begin{eqnarray}
u_{\mathrm{GS}}(r) &=&u_{0}r^{\sqrt{-U_{l}}}\exp \left( -\frac{1}{2}\sqrt{k}%
r^{2}\right) ,  \label{uGS} \\
\mu _{\mathrm{GS}} &=&\sqrt{k}\left( 1+\sqrt{-U_{l}}\right) ,  \label{muGS}
\end{eqnarray}%
with arbitrary constant $u_{0}$. An attempt to extend wave function (\ref%
{uGS}) to $U_{l}>0$ produces a formal wave function with asymptotic form $%
\sim \cos \left( \sqrt{U_{l}}\ln r\right) $ at $r\rightarrow 0$, which
actually does not make sense. The non-existence of the meaningful wave
function produced by the linearized equation (\ref{u}) for $U_{l}>0$ implies
the onset of the \textit{quantum collapse}, alias \textquotedblleft fall
onto the center" \cite{LL}, in the framework of the 2D linear Schr\"{o}%
dinger equation with the attractive potential (\ref{U}). This fact stresses
the crucial role of the nonlinear term in Eq. (\ref{u}), which makes it
possible to create the wave function with the meaningful asymptotic form (%
\ref{sigma=0}) for $U_{l}>0$.

As suggested in work \cite{we}, the consideration of singular solutions is
facilitated by the substitution of
\begin{equation}
u(r)\equiv r^{-2/\alpha }\chi (r),  \label{uchi}
\end{equation}%
which transforms Eq. (\ref{u}) into%
\begin{equation}
\mu \chi =-\frac{1}{2}\left[ \frac{d^{2}\chi }{dr^{2}}-\left( \frac{4}{%
\alpha }-1\right) \frac{1}{r}\frac{d\chi }{dr}+\frac{\left( U_{l}+4/\alpha
^{2}\right) }{r^{2}}\chi \right] +\frac{\chi ^{\alpha +1}}{r^{2}}.
\label{chi}
\end{equation}%
In Eq. (\ref{chi}) $k=0$ is set, as the effect of the trapping potential on
the singularity structure is negligible. Substitution (\ref{uchi}) separates
the singular factor, which is integrable (i.e., the respective norm
converges) and the singularity-free function $\chi (r)$, which takes a
finite value at $r=0$:%
\begin{equation}
\chi (r=0)=\left[ \frac{1}{2}\left( U_{l}+\frac{4}{\alpha ^{2}}\right) %
\right] ^{1/\alpha },  \label{r=0}
\end{equation}%
cf. Eq. (\ref{sigma=0}).

In what follows below, we present detailed results chiefly for the case of $%
\alpha =3$, which makes the asymptotic form (\ref{sigma=0}) more singular,
hence more interesting. The results for $\alpha =4$, i.e., the quintic
nonlinearity, which is relevant for the optical model, are presented below
too, in a more compact form.

\subsection{The stability problem}

\subsubsection{The Bogoliubov -- de Gennes equations}

The stability of stationary states can be addressed by taking perturbed
solutions as%
\begin{gather}
\psi (r,\theta ,t)=e^{-i\mu t+il\theta }r^{-2/\alpha }  \notag \\
\times \left[ \chi (r)+v_{1}(r)\exp \left( \Lambda t+im\theta \right)
+v_{2}^{\ast }(r)\exp \left( \Lambda ^{\ast }t-im\theta \right) \right] ,
\label{chi_perturbed}
\end{gather}%
where $\chi (r)$ is a solution of Eq. (\ref{chi}) (cf. Eq. (\ref{uchi})), $m$
is an integer angular index of small perturbations represented by the
eigenmode with components $v_{1,2}(r)$ (the asterisk stands for the complex
conjugate), and $\Lambda $ is the respective eigenvalue, which may be
complex. Instability takes place if there is at least one pair of
eigenvalues with Re$(\Lambda )\neq 0$.

The substitution of expression (\ref{chi_perturbed}) in Eq. (\ref{basic})
and linearization with respect to $v_{1,2}$ leads to the Bogoliubov - de
Gennes (BdG) equations \cite{Pit}. First, in the model with the quartic
nonlinearity ($\alpha =3$) they are
\begin{eqnarray}
\left( i\Lambda +\mu \right) v_{1}&=&-\frac{1}{2}\left[ \frac{d^{2}}{dr^{2}}-%
\frac{1}{3r}\frac{d}{dr}+\frac{U_{0}+4/9-(l+m)^{2}}{r^{2}}\right] v_{1}
\notag \\
&&+\frac{k}{2}r^{2}v_{1}+\frac{1}{2r^{2}}\chi ^{3}(r)(5v_{1}+3v_{2}),  \notag
\\
&&  \label{BdG} \\
\left( -i\Lambda +\mu \right) v_{2}&=&-\frac{1}{2}\left[ \frac{d^{2}}{dr^{2}}%
-\frac{1}{3r}\frac{d}{dr}+\frac{U_{0}+4/9-(l-m)^{2}}{r^{2}}\right] v_{2}
\notag \\
&&+\frac{k}{2}r^{2}v_{2}+\frac{1}{2r^{2}}\chi ^{3}(r)(5v_{2}+3v_{1}).  \notag
\end{eqnarray}%
Further, for the underlying quintic nonlinearity ($\alpha =4$) the BdG
equations are%
\begin{eqnarray}
\left( i\Lambda +\mu \right) v_{1}&=&-\frac{1}{2}\left[ \frac{d^{2}}{dr^{2}}+%
\frac{U_{0}+1/4-(l+m)^{2}}{r^{2}}\right] v_{1}  \notag \\
&&+\frac{k}{2}r^{2}v_{1}+\frac{1}{r^{2}}\chi ^{4}(r)(3v_{1}+2v_{2}),  \notag
\\
&&  \label{BdG2} \\
\left( -i\Lambda +\mu \right) v_{2}&=&-\frac{1}{2}\left[ \frac{d^{2}}{dr^{2}}%
+\frac{U_{0}+1/4-(l-m)^{2}}{r^{2}}\right] v_{2}  \notag \\
&&+\frac{k}{2}r^{2}v_{2}+\frac{1}{r^{2}}\chi ^{4}(r)(3v_{2}+2v_{1}).  \notag
\end{eqnarray}

For the special flat-density solution given by Eq. (\ref{flat}) (i.e., and $%
k=0$ and $U_{0}=l^{2}$), the perturbed solution is introduced as (cf. Eq. (%
\ref{chi_perturbed}))
\begin{gather}
\psi (r,\theta ,t)=e^{-i\mu t+il\theta }  \notag \\
\times \left[ \mu ^{1/\alpha }+v_{1}(r)\exp \left( \Lambda t+im\theta
\right) +v_{2}^{\ast }(r)\exp \left( \Lambda ^{\ast }t-im\theta \right) %
\right] ,  \label{pert-special}
\end{gather}%
and the BdG equations take a simpler form,%
\begin{eqnarray}
i\Lambda v_{1} &=&-\frac{1}{2}\left( \frac{d^{2}}{dr^{2}}+\frac{1}{r}\frac{d%
}{dr}-\frac{m^{2}+2lm}{r^{2}}\right) v_{1}  \notag \\
&&+\beta \mu (v_{1}+v_{2}),  \notag \\
&&  \label{special} \\
-i\Lambda v_{2} &=&-\frac{1}{2}\left( \frac{d^{2}}{dr^{2}}+\frac{1}{r}\frac{d%
}{dr}-\frac{m^{2}-2lm}{r^{2}}\right) v_{2}  \notag \\
&&+\beta \mu (v_{2}+v_{1}),  \notag
\end{eqnarray}%
where $\beta \left( \alpha =3\right) \equiv 3/2$ and $\beta (\alpha
=4)\equiv 2$.

\subsubsection{Analytical results for the stability}

While BdG equations (\ref{BdG}), (\ref{BdG2}) and (\ref{special}) should be
solved numerically, partial results for the stability can be obtained in an
analytical form, in spite of the apparent complexity of the equations. In
particular, for the singular vortex states (with $l\geq 1$) an asymptotic
analysis of Eq. (\ref{BdG}) at $r\rightarrow 0$ was performed in work \cite%
{we}, looking for solutions as%
\begin{equation}
v_{1, 2}(r)\approx v_{1, 2}^{(0)}r^{\gamma },  \label{gamma}
\end{equation}%
with $\gamma $ determined by equations%
\begin{equation}
\gamma ^{2}-\frac{4}{3}\gamma -3\chi ^{3}(r=0)-m^{2}=\pm \sqrt{%
4l^{2}m^{2}+9\chi ^{6}(r=0)},  \label{gamma2}
\end{equation}%
where $\chi ^{3}(r=0)$ is taken as per Eq. (\ref{r=0}). Relevant solutions
of Eq. (\ref{gamma2}) are ones with Re$\left( \gamma \right) \geq 0$
(otherwise, eigenmode (\ref{gamma}) is inappropriate). The emergence of an
eigenmode which may bring instability is signalized by a solution of Eq. (%
\ref{gamma2}) crossing $\gamma =0$. The critical role is played by the modes
with $m=\pm 1$, which account for drift instability of the vortex (the onset
of spontaneous motion of the vortex' pivot along a spiral trajectory,
drifting away from $r=0$, cf. Figs. \ref{fig2}(c) and \ref{fig7} (c,d)
presented below; in certain cases, usual (regular) dark vortices may be
subject to a similar instability \cite{Rokhsar}). Thus, a straightforward
consideration of Eq. (\ref{gamma2}) with $m^{2}=1$ readily demonstrates the
singular vortex may be stable, as a solution to Eq. (\ref{basic}) with $%
\alpha =3$, in the region of
\begin{equation}
U_{0}\geq \left( U_{0}\right) _{\min }=\left( 7/9\right) \left(
3l^{2}-1\right) ,  \label{as_before}
\end{equation}%
where the drift-perturbation mode does not exist. Because the asymptotic
consideration at $r\rightarrow 0$, which leads to Eq. (\ref{as_before}), is
not altered in the presence of the background far from $r=0$, condition (\ref%
{as_before}) is equally relevant for the singular vortices in the present
case. Numerical results, presented in the next section, confirm that this
condition accurately determines the stability of the vortices with $l\geq 1$.

A similar analysis can be performed using the BdG equations (\ref{BdG2})\
for singular vortices produced by Eq. (\ref{basic}) with $\alpha =4$ (the
quintic nonlinearity). In this case, the substitution of ansatz (\ref{gamma}%
) for the asymptotic eigenmodes leads to the following equation, instead of
Eq. (\ref{gamma2}):%
\begin{equation}
\gamma ^{2}-\gamma -4\chi ^{4}(r=0)-m^{2}=\pm 2\sqrt{l^{2}m^{2}+4\chi
^{8}(r=0)},  \label{gamma3}
\end{equation}%
where $\chi \left( r=0\right) $ should again be taken from Eq. (\ref{r=0}).
The drifting instability, corresponding to $m^{2}=1$, is absent in the
region of%
\begin{equation}
U_{0}\geq 2l^{2}-1/2,  \label{gamma4}
\end{equation}%
the boundary of which is also defined as the zero-crossing of $\gamma $. It
is worthy to note the similarity of the stability conditions (\ref{as_before}%
) and (\ref{gamma4}).

For the special flat-vortex state (\ref{flat}) numerical solution of the BdG
equations (\ref{special}) around this state is a challenging issue, as the
result is not stable enough against variation of technical details, such as
the mesh size of the underlying numerical scheme. On the other hand, there
is a simple argument in favor of stability of the flat vortex. Indeed, it is
easy to find an asymptotic form of eigenmodes produced by Eq. (\ref{special}%
) at $r\rightarrow 0$, choosing $m>0$, for the sake of the definiteness:%
\begin{equation}
v_{2}\approx \mathrm{const}\cdot r^{\sqrt{m^{2}-2ml}},~v_{1}\approx \frac{%
\beta \mu }{2}\frac{\mathrm{const}\cdot r^{\sqrt{m^{2}-2ml}+2}}{\sqrt{%
m^{2}-2lm}+1-lm}.  \label{v21}
\end{equation}%
The eigenmodes given by Eq. (\ref{v21}) are relevant ones for $m\geq 2l$,
otherwise they produce singular expressions. The latter condition excludes $%
m=1$, hence the above-mentioned drift instability in ruled out. Further,
usual dark vortices with $l\geq 2$ tend to be unstable against spontaneous
splitting into a set of $l$ unitary eddies, as demonstrated in detail in
various settings \cite{HanPu, Neu, Kawaguchi, Finland, Delgado, Poland} (see
also Fig. \ref{fig13} below). Because a growing perturbation eigenmode with
azimuthal index $m\geq 2$ splits the multiple vortex into a set of $m$
fragments, the perturbation azimuthal index which might be responsible for
the splitting instability is $m=l$, hence the condition $m\geq 2l$ excludes
this instability as well.

\subsection{The Thomas-Fermi (TF) approximation}

The TF approximation may be applied either to Eq. (\ref{chi}) by dropping
derivatives in it, or similarly to Eq. (\ref{u}). In the former case, the
result is$\allowbreak $%
\begin{equation}
u_{\mathrm{TF}}(r)\equiv r^{-2/3}\chi _{\mathrm{TF}}(r)=\left( \mu +\frac{1}{%
2}\left( U_{l}+\frac{4}{\alpha ^{2}}\right) r^{-2}\right) ^{1/\alpha }.
\label{TF1}
\end{equation}%
For $U_{l}+4/\alpha ^{2}>0$, this approximation is valid at all values of $r$%
, assuming that $k=0$ is set in Eq. (\ref{u}). In particular, it yields the
correct singular (but physically acceptable) form of the solution at $%
r\rightarrow 0$, which is fully tantamount to Eq. (\ref{singular}). If the
TF approximation is applied directly to Eq. (\ref{u}) with $k=0$, the result
is different:%
\begin{equation}
\tilde{u}_{\mathrm{TF}}(r)=\left( \mu +\frac{U_{l}}{2}r^{-2}\right)
^{1/\alpha }.  \label{tilde}
\end{equation}%
While TF approximation (\ref{TF1}) is more accurate at $r\rightarrow 0$, the
expansion of the alternative approximation (\ref{tilde}) with $k=0$ at $%
r\rightarrow \infty $ yields a result which is fully tantamount to the
correct asymptotic form (\ref{large r}).

A nontrivial issue which is suggested by the combination of both versions of
the TF approximation is a possibility of the existence of the antidark
singular modes in interval (\ref{4/9}), where Eq. (\ref{TF1}) predicts a
profile which is a monotonously decaying function of $r$, while both Eqs. (%
\ref{tilde}) and (\ref{large r}) demonstrate that, at $r\rightarrow \infty $%
, $u(r)$ approaches the background value $\mu ^{1/3}$ \emph{from below}.
These conflicting predictions suggest that, in interval (\ref{4/9}), there
may exist a mode with a \textit{non-monotonous} radial profile, i.e., with a
local minimum of $u(r)$ at some finite $r$. Numerical results displayed
below confirm this conjecture (see Fig. \ref{fig3}(c)).

For regular dark vortices existing at $U_{l}<0$, with $u(r)$ vanishing at $%
r\rightarrow 0$, substitution (\ref{uchi}) is irrelevant, therefore the
appropriate TF approximation is the one applied to Eq. (\ref{u}). It
predicts a structure which, as usual, includes artifacts in the form of an
\textit{inner hole} at small $r$ and \textit{zero tail} at large $r$ (cf.
\cite{BEC-review3}):%
\begin{equation}
\tilde{u}_{\mathrm{TF}}(r)=\left\{
\begin{array}{c}
0,~\mathrm{at}~r^{2}>r_{+}^{2}\equiv k^{-1}\left( \mu +\sqrt{\mu
^{2}-k\left\vert U_{l}\right\vert }\right) ,~ \\
\left[ \mu -(\left\vert U_{l}\right\vert /2)r^{-2}-\left( k/2\right) r^{2}%
\right] ^{1/\alpha },~\mathrm{at}~r_{-}^{2}<r^{2}<r_{+}^{2}, \\
0,~\mathrm{at}~r^{2}<r_{-}^{2}\equiv k^{-1}\left( \mu -\sqrt{\mu
^{2}-k\left\vert U_{l}\right\vert }\right) .%
\end{array}%
\right.  \label{TF2}
\end{equation}%
This TF solution exists above the threshold,
\begin{equation}
\mu >\mu _{\mathrm{thr}}\equiv \sqrt{k\left\vert U_{l}\right\vert },
\label{mumin}
\end{equation}%
which is the TF limit of the ground-state eigenvalue (\ref{muGS}) produced
by the linearized version of Eq. (\ref{u}). Equation (\ref{TF2}) predicts a
maximum of the TF field at
\begin{equation}
\left( r_{\max }\right) _{\mathrm{TF}}=\left( \left\vert U_{l}\right\vert
/k\right) ^{1/4}.  \label{rmax}
\end{equation}%
Note that, in the TF approximation, expressions (\ref{mumin}) and (\ref{rmax}%
) do not depend on $\alpha $.

In the case of $k=0$, the above-mentioned zero-tail artifact disappears. In
this case, $\mu _{\mathrm{thr}}$, as given by Eq. (\ref{mumin}), vanishes
too, and the TF approximation simplifies to
\begin{equation}
u_{\mathrm{TF}}(r)=\left\{
\begin{array}{c}
\left[ \mu -(\left\vert U_{l}\right\vert /2)r^{-2}\right] ^{1/\alpha },~%
\mathrm{at}~r^{2}>r_{-}^{2}(k=0)\equiv \left\vert U_{l}\right\vert /\left(
2\mu \right) , \\
0,~\mathrm{at}~r^{2}<\left\vert U_{l}\right\vert /\left( 2\mu \right) .%
\end{array}%
\right.  \label{TF3}
\end{equation}%
Note that the TF solution (\ref{TF3}) agrees with the correct asymptotic
form (\ref{large r}) at $r\rightarrow \infty $

True dark-vortex solutions include no hole at small $r$. Instead, the
correct asymptotic form at $r\rightarrow 0$ is
\begin{equation}
u(r)\approx \mathrm{const}\cdot r^{\sqrt{-U_{l}}},  \label{dark}
\end{equation}%
as per Eq. (\ref{uGS}). In the case of $U_{0}=0$, i.e., $U_{l}=-l^{2}$ (see
Eq. (\ref{Ul})), this asymptotic form carries over into the usual one for
the dark vortices, given by Eq. (\ref{r^l}). Note also that the regular dark
vortices exist in the presence of the attractive potential ($U_{0}>0$),
provided that it is not too strong, $U_{0}<l^{2}$, making $U_{l}$ negative.



\section{Numerical results}

Stationary solutions of Eq. (\ref{basic}) were obtained by dint of the
Newton's iteration method, and their stability was identified through
numerical solution of BdG equations (\ref{BdG}). Then, the predicted
(in)stability was verified in simulations of Eq. (\ref{basic}) for perturbed
evolution of the solutions in question. The simulations were carried out in
the Cartesian coordinates, using the split-step fast-Fourier-transform
method. The numerical results are presented below chiefly for the quartic
nonlinearity ($\alpha =3$), because, as mentioned above, the solutions are
more singular in this case, make it more interesting. For the quintic
nonlinearity ($\alpha =4$), numerical results, which are not shown here in
detail, are quite similar.

\subsection{Singular and intermediate states}

\begin{figure}[tbph]
\centering \includegraphics[width=17cm]{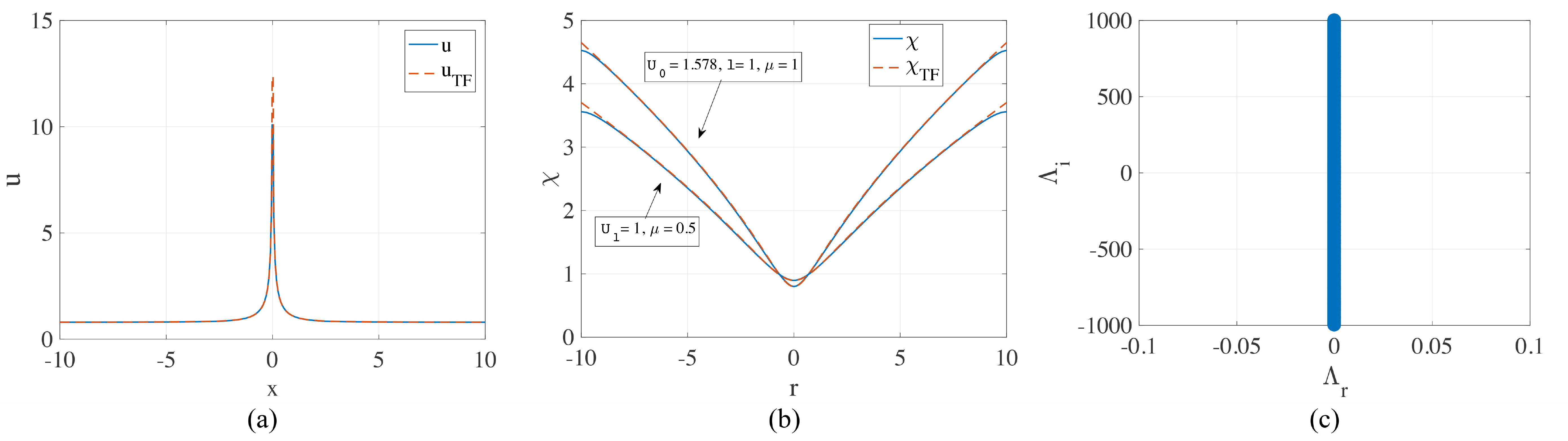}
\caption{(a) The profile of the singular state, produced by the numerical
solution of Eq. (\protect\ref{u}) with $\protect\alpha =3$ and $k=0$, and
its counterpart, predicted by the TF approximation (\protect\ref{TF1}), for $%
U_{l}=1$ and $\protect\mu =0.5$. (b) The lower curves are the same as (a),
shown in terms of function $\protect\chi (r)$, see Eq. (\protect\ref{uchi}).
The upper curves in (b) show an example of a stable singular vortex, taken
at values of parameters close to the stability boundary given by Eq. (%
\protect\ref{as_before}): $l=1$, $\protect\mu =1$, and $U_{0}=1.578$, while
the boundary value is $\left( U_{0}\right) _{\min }=1.556$. (c) The spectrum
of the stability eigenvalues for the singular vortex corresponding to the
upper plots in (a), as produced by the numerical solution of BdG equations (%
\protect\ref{BdG}).}
\label{fig1}
\end{figure}

At all values of the parameters, i.e., $\mu >0$, $U_{0}>0$ (interval (\ref%
{4/9}) is considered separately below), and $l\geq 0$, the numerically found
profiles of the antidark singular states are virtually identical to their TF
counterparts given by Eq. (\ref{TF1}). The same is true for the quintic
nonlinearity, with $\alpha =4$ in Eq. (\ref{u}) (not shown here in detail).
A typical example is displayed in Fig. \ref{fig1} for $U_{l}=1$ (see Eq. (%
\ref{Ul})) and $\mu =0.5$. While panel \ref{fig1}(a), which shows the global
shape of the singular solution, does not make it possible to discern details
of the singular peak, this is shown by the lower lines in panel \ref{fig1}%
(b), by means of function $\chi (r)$, from which the singular factor is
eliminated by means of substitution (\ref{uchi}).

\begin{figure}[tbph]
\centering\includegraphics[width=16cm]{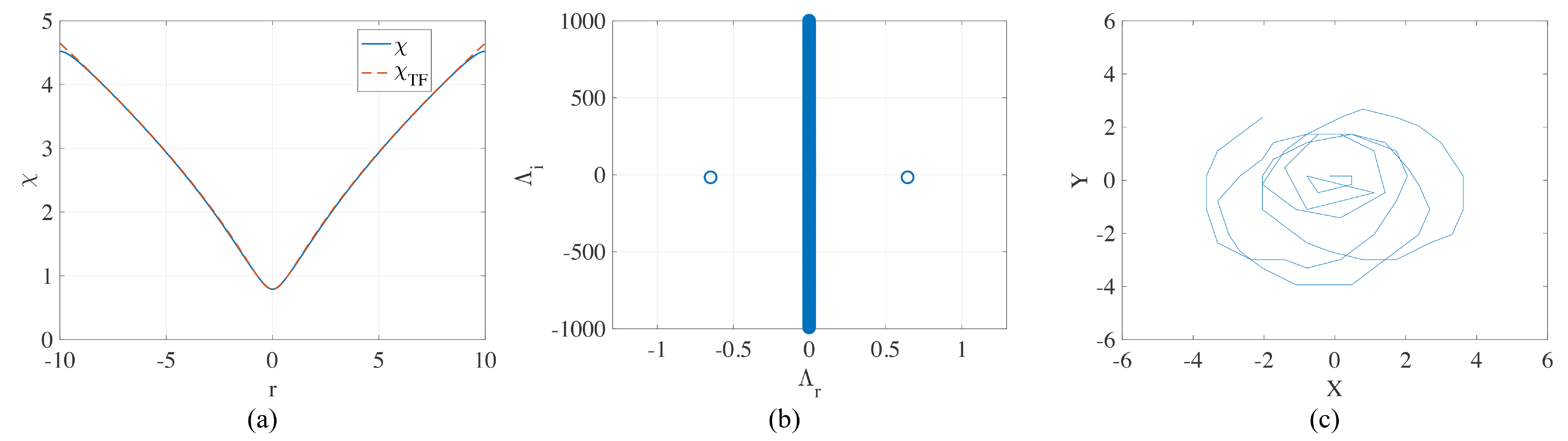}
\caption{(a,b) The same as in Figs. \protect\ref{fig1}(b,c), but for $%
U_{0}=1.533$. This state is unstable, in agreement with Eq. (\protect\ref%
{as_before}). Panel (c) shows the trajectory of spontaneous motion of the
vortex' pivot, produced by simulations of Eq. (\protect\ref{basic}) for
total evolution time $t=300$. }
\label{fig2}
\end{figure}

Further, up to the numerical accuracy, the results obtained for the
stability of the singular states exactly agree with the analytical stability
condition given by Eq. (\ref{as_before}). This conclusion is illustrated by
Figs. \ref{fig1}(b,c) and \ref{fig2}, where the singular vortices with $l=1$%
, $\mu =1$ and spectra of their stability eigenvalues are displayed,
respectively, for $U_{0}=1.578$ and $U_{0}=1.533$, while the respective
stability-boundary value is $\left( U_{0}\right) _{\min }=14/9\approx
\allowbreak 1.556$. It is seen that, accordingly, the former solution is
stable, while the latter one is not. Additionally, Fig. \ref{fig2}(c)
displays the trajectory of spontaneous motion of the pivot caused by the
drift instability of the stationary vortex shown in Fig. \ref{fig2}(a). The
drift character of the instability complies with the fact that the unstable
eigenvalues, shown in panel \ref{fig2}(b), are produced by small
perturbations with $m=1$ in Eq. (\ref{chi_perturbed}).

\begin{figure}[tbph]
\centering\includegraphics[width=16cm]{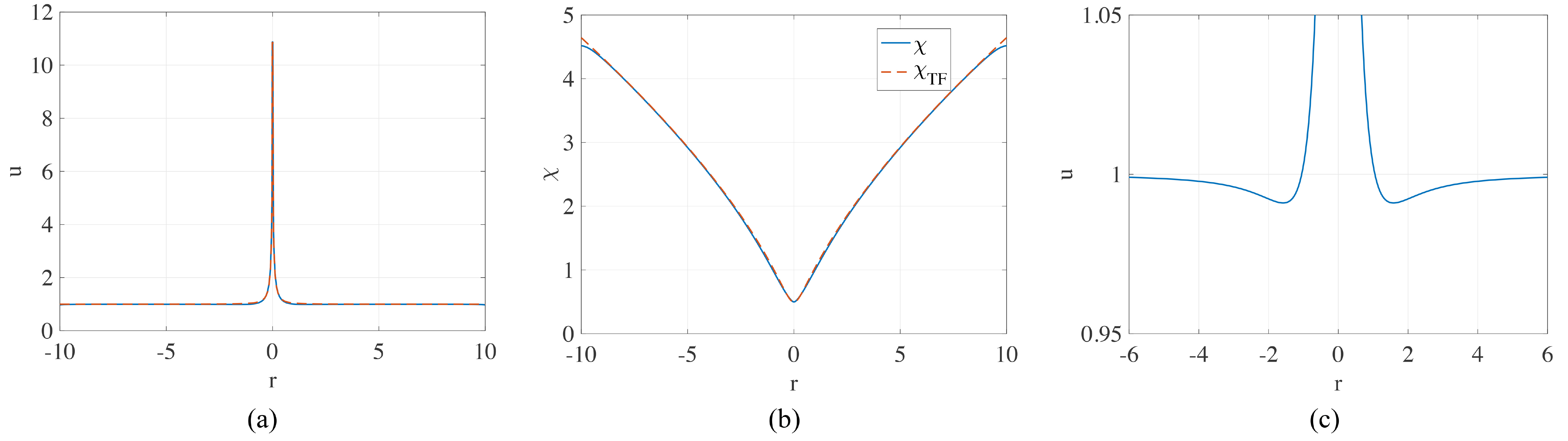}
\caption{Panels (a) and (b): the same as in Figs. \protect\ref{fig1}(a,b),
but for $l=0$, $\protect\mu =1$, and $U_{0}=-0.2$ (this value belongs to
interval (\protect\ref{4/9})). Panel (c) shows a shallow local minimum in
the numerically found radial profile of the solution. This state is stable,
as are all solutions with $l=0$ belonging to interval (\protect\ref{4/9}).}
\label{fig3}
\end{figure}

Numerical results confirm as well the existence of the \textquotedblleft
counter-intuitive" singular antidark modes in interval (\ref{4/9}). An
example, and its comparison to the TF approximation, as given by Eq. (\ref%
{TF1}), are displayed in Fig. \ref{fig3}. In particular, the zoom of a
segment of the modal profile shown in panel (c) clearly confirms the
existence of a shallow local minimum of $u(r)$. As mentioned above, this
feature is made necessary by the impossibility to directly match the
asymptotic form of the singularity, given by Eq. (\ref{sigma=0}), and the
asymptotic expansion (\ref{large r}) at $r\rightarrow \infty $. The singular
states with $l=0$ are stable in interval (\ref{4/9}), while their
counterparts with $l\geq 1$ are not, in accordance with Eq. (\ref{as_before}%
).

\begin{figure}[tbph]
\centering\includegraphics[width=8cm]{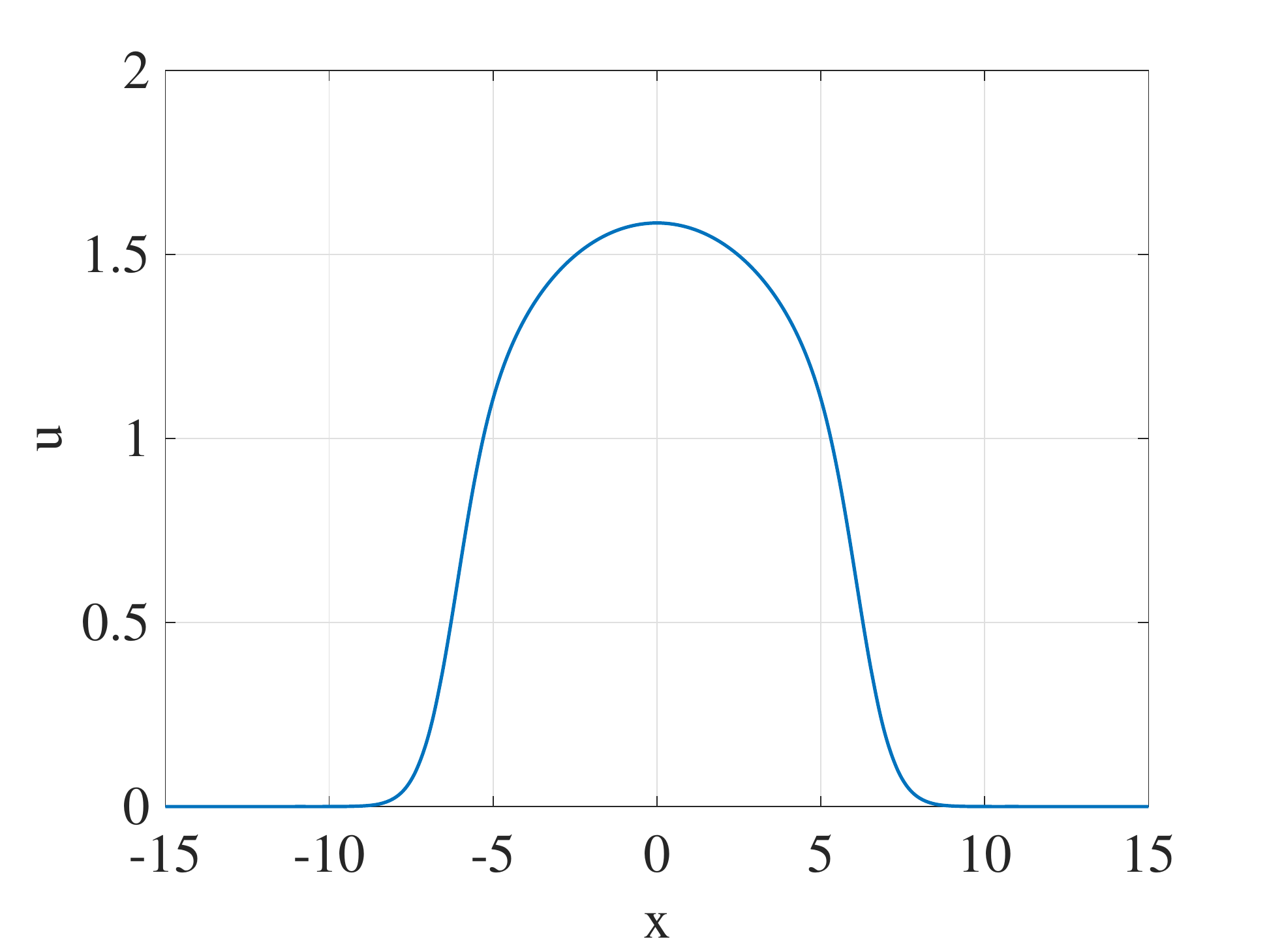}
\caption{The numerically found profile of the intermediate vortex model with
$U_{l}=0$ (the one which has the flat profile (\protect\ref{flat}) in the
absence of the HO potential), for $\protect\mu =4$, $k=0.2$, and $\protect%
\alpha =3$.}
\label{fig4}
\end{figure}

The above singular states are produced by setting $k=0$ in Eqs. (\ref{basic}%
) and (\ref{u}), as the HO trapping potential does not play a significant
role for the singular states. 
On the other hand, the HO potential produces an essential effect if applied
to the flat-vortex mode (\ref{flat}), since it distorts the flat profile, as
shown in Fig. \ref{fig4}. Note that the width of the profile is close to
that of the GS wave function of the linearized equation, given by Eq. (\ref%
{uGS}), while its amplitude is accurately predicted by the TF approximation,
which yields $u_{\mathrm{TF}}(x=0)=\mu ^{1/3}\approx 1.59$ in this case.

\subsection{Excitation of singular vortices by regular vortex inputs}

An issue which is crucially important for the possibility of the creation of
singular (antidark) vortices in the experiment, using the nonlinear light
propagation and BEC alike, is a scenario for the excitation of such
\textquotedblleft extraordinary" modes, as incident optical beams, which are
employed for the creation of the usual photonic \cite{Swartz}-\cite%
{opt-review3}, \cite{Neshev,Segev} and matter-wave \cite%
{vort-excit1,vort-excit2}, \cite{BEC-review1}-\cite{BEC-review3} vortices,
may carry only the ordinary (nonsingular) vorticity.

Thus, it is relevant to simulate Eq. (\ref{basic}) with $U_{0}>0$ and the
input taken as a usual vortex beam, i.e., a stationary solution of the
linearized version of Eq. (\ref{basic}) with $U_{0}=0$:%
\begin{equation}
\psi _{0}\left( r,\theta \right) =a_{0}r^{l}e^{il\theta }\exp \left( -\frac{1%
}{2}\sqrt{k}r^{2}\right) ,  \label{psi0}
\end{equation}%
where $a_{0}$ is an arbitrary constant. To the best of our knowledge, such
simulations were not reported in previous works. Note that input (\ref{psi0}%
) cannot generate an ordinary (dark nonsingular) vortex state, as it does
not exist in the case of $U_{0}>0$, according to Eq. (\ref{dark}).

\begin{figure}[tbph]
\centering\includegraphics[width=16cm]{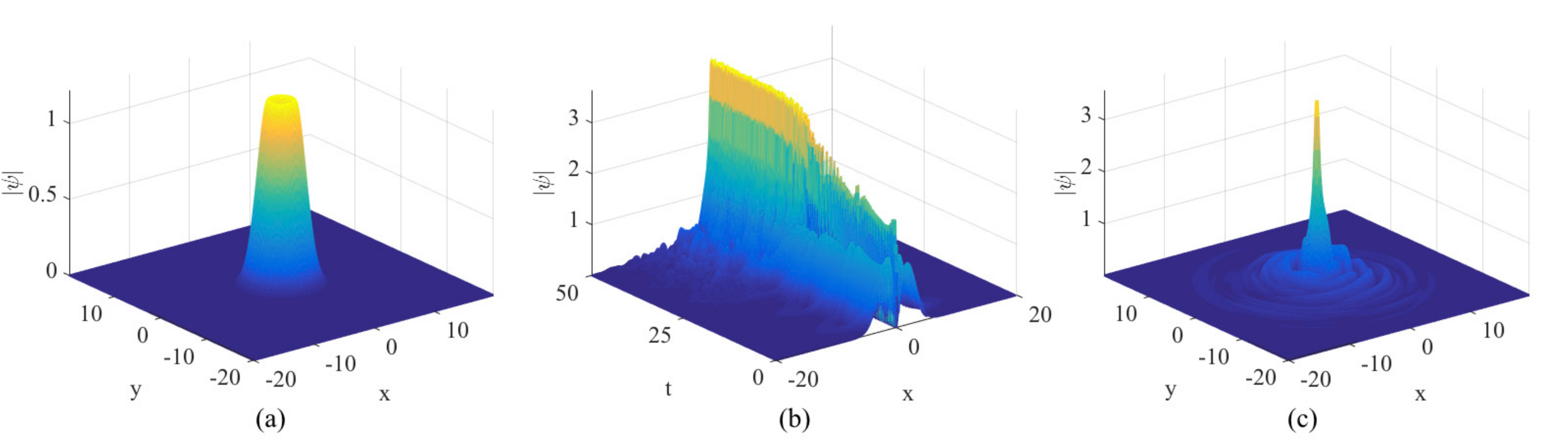}
\caption{Creation of a singular vortex in simulations of Eq. (\protect\ref%
{basic}), with $U_{0}=4$, $k=0.1$, and $\protect\alpha =3$, from the
regular-vortex input taken as per Eq. (\protect\ref{psi0}) with $l=1$ and
total norm $N=40$ (see Eq. (\protect\ref{N})), which corresponds to $a_{0}=2/%
\protect\sqrt{\protect\pi }\approx 1.13$ in Eq. (\protect\ref{psi0}). Panels
(a) and (c): amplitude profiles of the input (at $t=0$) and output (at $t=50$%
), respectively. The intermediate panel (b) displays the spatiotemporal
evolution of $\left\vert \protect\psi \left( x,t\right) \right\vert $ in
cross section $y=0$.}
\label{excitation}
\end{figure}

Direct simulations of Eq. (\ref{basic}) readily confirm that, at all values $%
U_{0}>0$, and for both nonlinearities considered here, which correspond to $%
\alpha =3$ and $4$, initial condition (\ref{psi0}) gives rise to perturbed
vortex states, close to the expected singular ones, which keep the initial
vorticity, $l$. 
A typical example is presented in Fig. \ref{excitation}, which displays
amplitude profiles of the input and output vortex modes (panels (a) and (c),
respectively) and the overall image of the spatiotemporal evolution from the
regular initial condition to a quasi-singular output, shown by means of the
radial cross section in panel (b).

In Fig. \ref{excitation}(c), it is observed that the emerging central
singular vortex (in the simulation, its amplitude remains finite due to the
effect of the Cartesian numerical mesh) is surrounded by an additional set
of radial waves. The central core and radial undulations are separated by a
zero-amplitude ring, whose radius may be predicted by the TF approximation,
applied to Eq. (\ref{u}) with $\mu =0$. Indeed, such an approximation
predicts $u(r)=0$ at $r_{0}=\left( U_{l}/k\right) ^{1/4}$. For parameters of
Fig. \ref{excitation}, this formula yields $r_{0}\approx 2.34$, which is
close to the zero-amplitude radius observed in Fig. \ref{excitation}(c).

The fact that the simulations displayed in Fig. \ref{excitation} were
performed in the Cartesian coordinates explains some deviation from the
axial symmetry observed in the figure (in particular, the pivot of the
emerging singular vortex is somewhat shifted off the central point and
performs circular motion around it). The simulations used an absorber
installed at edges of the spatial domain. As a result, about $2\%$ of the
initial norm was lost in the course of the simulations, by $t=50$.

\subsection{Regular dark vortices}

We address the regular dark modes, first, in the case of the repulsive
central potential, i.e., $U_{0}<0$, with $l=0$ and $1$, supported by the
flat background (\ref{large r}) in the absence of the HO trapping potential,
$k=0$. Different values of $U_{0}$ and $l$, which amount to the same value
of combination (\ref{Ul}), give rise to identical shapes of the modes (while
their stability depends on $U_{l}$ and $l$ separately, as shown below). In
Fig. \ref{fig5} we plot a set of profiles of the dark modes produced by the
numerical solution of Eq. (\ref{u}) for a fixed value of the chemical
potential, $\mu =0.5$, and three different values of $U_{l}$, \textit{viz}.,
$-1$, $-2$, and $-3$. In the same figure, the numerically found profiles are
compared to the TF approximation, as given by Eq. (\ref{TF3}).
\begin{figure}[tbph]
\centering\includegraphics[width=16cm]{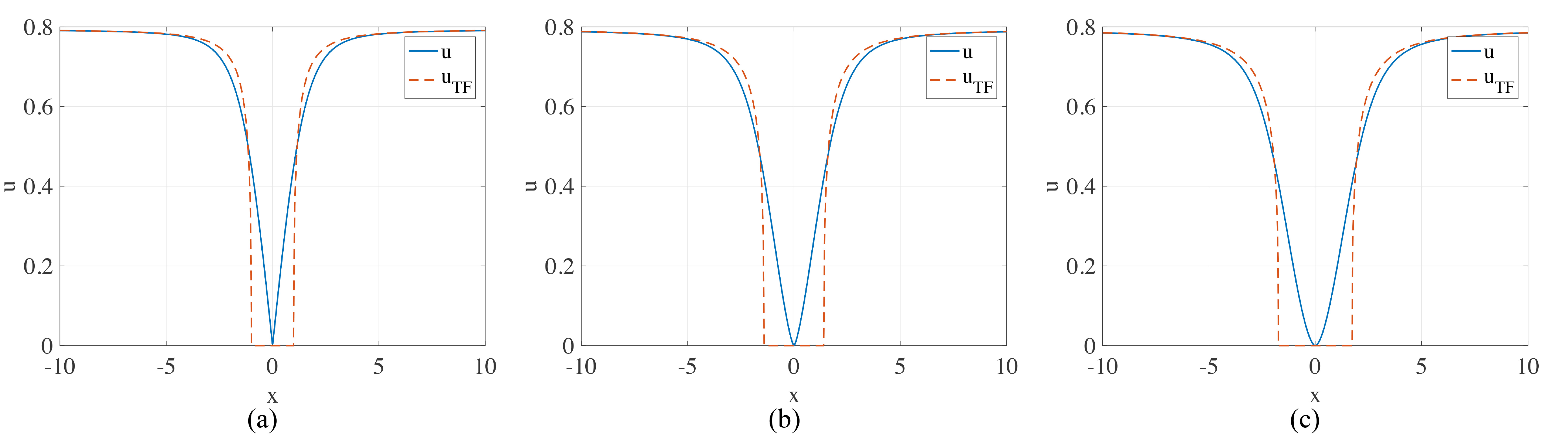}
\caption{Shapes of dark regular modes produced by the numerical solution of
Eq. (\protect\ref{u}) with $k=0$ (no HO trap), $\protect\alpha =3$, $\protect%
\mu =0.5$ and $U_{l}=-1$ (a), $-2$ (b), and $-3$ (c). Their counterparts
produced by the TF approximation based on Eq. (\protect\ref{TF3}) are shown
too.}
\label{fig5}
\end{figure}

\begin{figure}[tbph]
\centering\includegraphics[width=16cm]{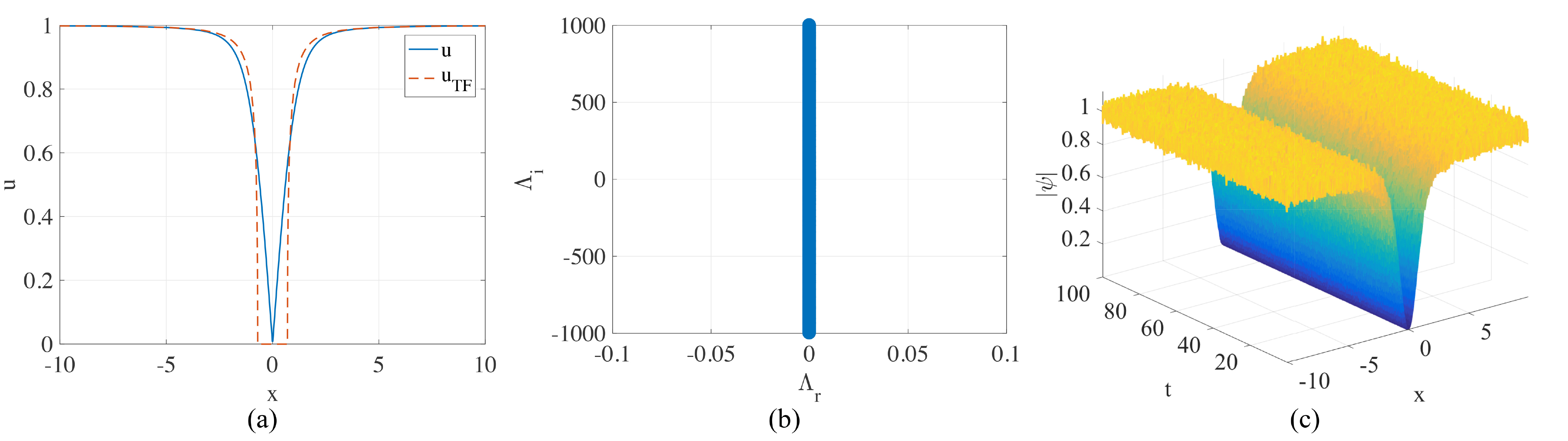}
\caption{(a) The numerically found shape of the dark mode, obtained as a
solution of Eq. (\protect\ref{u}) with $k=0$ (no HO\ trapping potential), $%
\protect\alpha =3$, $U_{0}=-1$, $l=0$, and $\protect\mu =1$. Its TF
counterpart is displayed too, as produced by Eq. (\protect\ref{TF3}). (b)
The spectrum of stability eigenvalues for small perturbations around this
mode, $\Lambda \equiv \Lambda _{\mathrm{r}}+i\Lambda _{\mathrm{i}}$, as
produced by the numerical solution of the respective BdG equations. The
spectrum includes no unstable eigenvalues with $\Lambda _{\mathrm{r}}\neq 0$%
. (c) The simulated evolution of the same mode, with $5\%$ initial random
perturbations added to it, confirms its stability.}
\label{fig6}
\end{figure}

Both the calculation of the stability eigenvalues through the numerical
solution of Eq. (\ref{BdG}) and direct simulations demonstrate that all the
dark modes with $l=0$ and $1$ are stable as solutions of Eq. (\ref{basic})
with $U_{0}<0$ and $k=0$. An example is presented in Fig. \ref{fig6}, which
demonstrates the shape, spectrum of eigenvalues, and perturbed evolution of
a typical solution of such a type.

As mentioned above, the dark vortex with $l\geq 1$ may exist in the case of
the attractive central potential, \textit{viz}., at%
\begin{equation}
0<U_{0}<l^{2},  \label{01}
\end{equation}%
which corresponds to $0<-U_{l}<l^{2}$, as per Eq. (\ref{Ul}). Even for the
simplest case of $l=1$, results for the stability are complicated in this
case. The numerical analysis performed for $l=1$ reveals an alternation of
stability and instability regions in the interval of $0<U_{0}<1$. The
analysis was performed with the inclusion of a weak but finite trapping HO\
potential in Eqs. (\ref{basic}), (\ref{u}), and the corresponding BdG
equations, as, in the absence of the trap, the necessity to maintain the
vortical boundary conditions at $r\rightarrow \infty $ makes it difficult to
solve the stability problem, cf. work \cite{HanPu}. As an example, Fig. \ref%
{fig7} presents the shape and spectrum of the stability eigenvalues for the
dark vortex, as obtained at a small positive value of the attraction
strength, $U_{0}=0.03$, with $\mu =1$ and a weak trapping potential,
corresponding to $k=0.02$ in Eq. (\ref{basic}). 
\begin{figure}[tbph]
\centering\includegraphics[width=16cm]{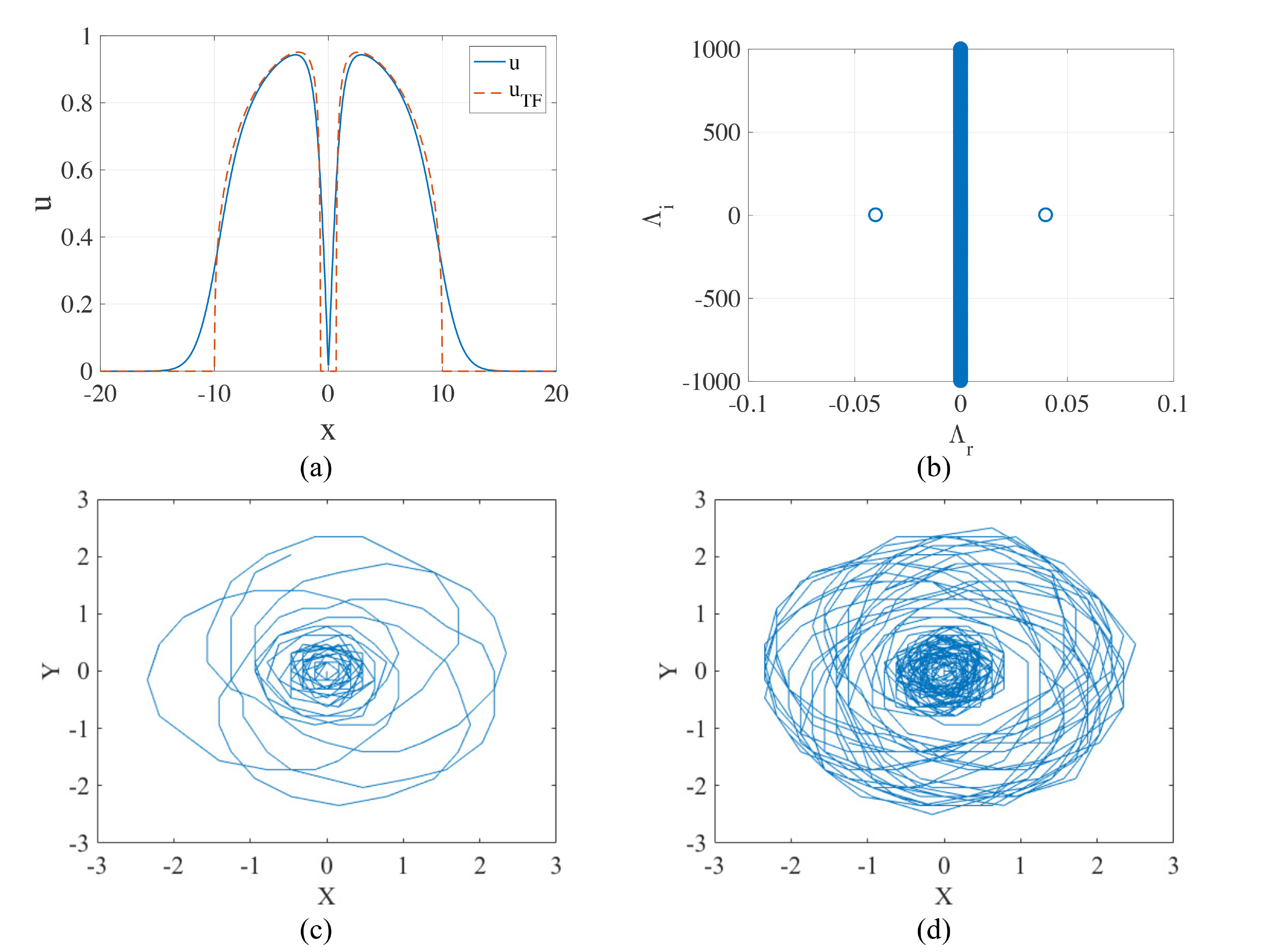}
\caption{Panels (a) and (b) show the same as (a) and (b) in Fig. \protect\ref%
{fig6}, but for an unstable dark vortex, produced by the numerical solution
of Eqs. (\protect\ref{u}) and (\protect\ref{BdG}) with $U_{0}=0.03$, $k=0.02$%
, for $l=1$ and $\protect\mu =1$. The TF profile in (a) is produced by Eqs. (%
\protect\ref{TF2}) and (\protect\ref{rmax}). The unstable pair of
eigenvalues in (b) corresponds to the azimuthal perturbation index $m=1$,
cf. Eq. (\protect\ref{chi_perturbed}). Trajectories of the motion of the
central point (pivot) of an unstable dark vortex from (a), as produced by
simulations of Eq. (\protect\ref{basic}) with total evolution time $t=1000$
and $4000$ are displayed in (c) and (d), respectively.}
\label{fig7}
\end{figure}

The instability represented by the pair of real eigenvalues in Fig. \ref%
{fig7}(b) corresponds to azimuthal index $m=1$ of small perturbations, cf.
Eq. (\ref{chi_perturbed}), which, as said above, implies the drift
instability. In accordance with the expectation, in direct simulations the
pivot of the unstable dark vortex spontaneously starts motion along a
spiral-like unwinding trajectory, which however, remains trapped in the area
of $r<r_{\mathrm{trap}}=k^{-1/4}\approx 2.7$, where $r_{\mathrm{trap}}$ is
the trapping radius imposed by the HO potential, see Eq. (\ref{uGS}). Motion
of the pivot along the confined trajectory of an apparently irregular shape
was going on as long as the simulations were running.

\begin{figure}[tbph]
\centering\subfigure[]{\includegraphics[width=8cm]{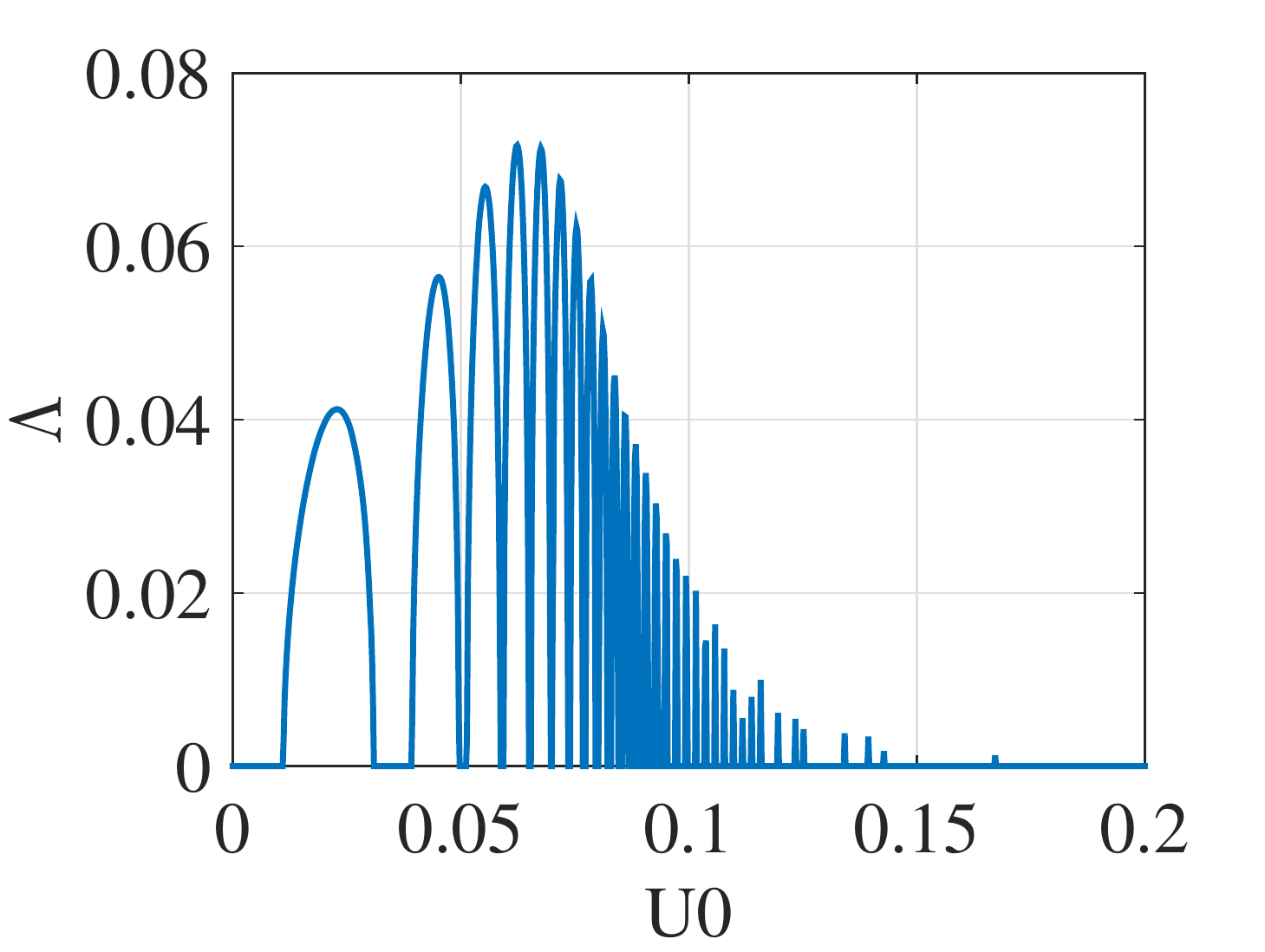}} %
\subfigure[]{\includegraphics[width=8cm]{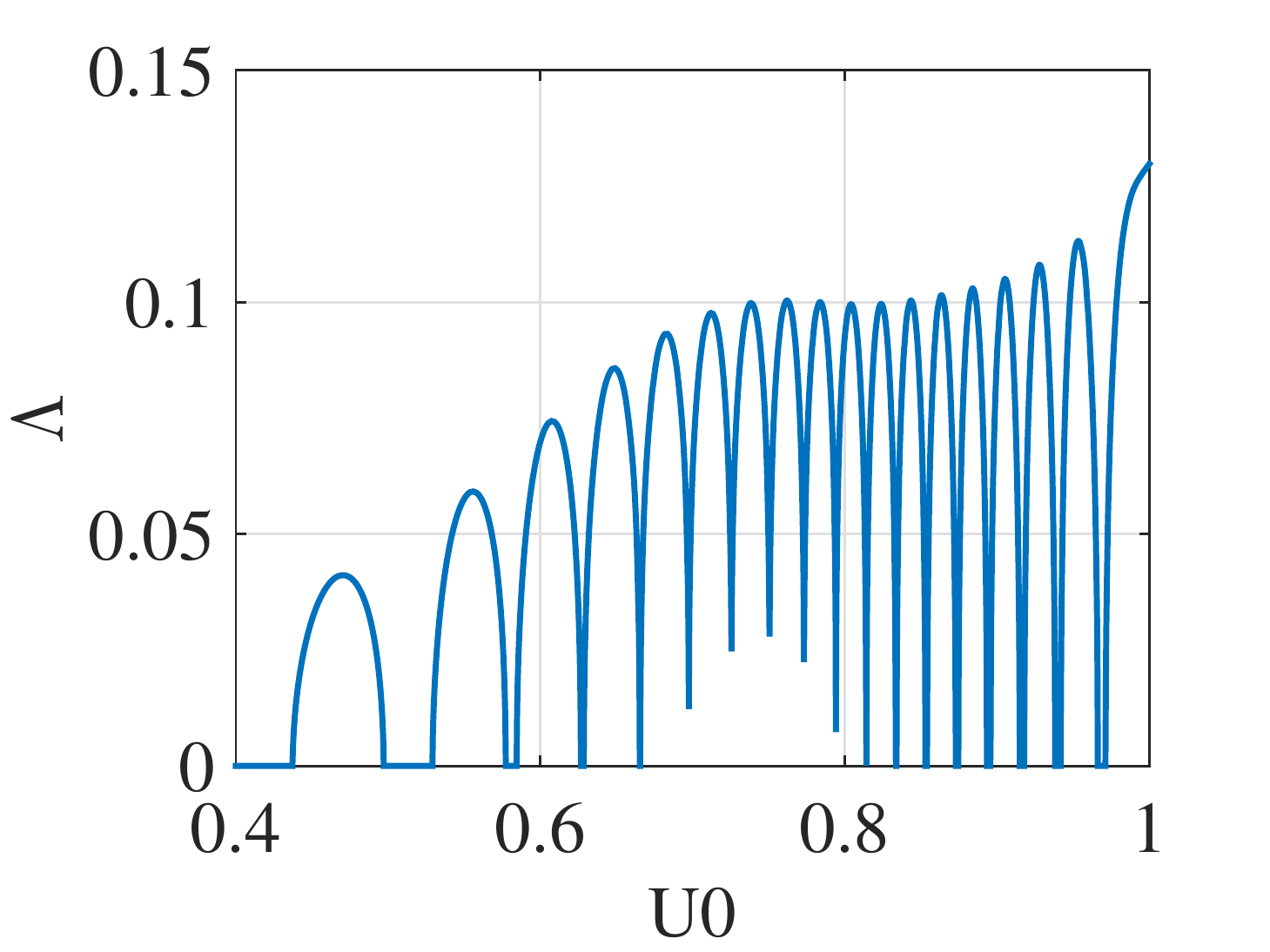}}
\caption{The real instability growth rate, $\Lambda $, for the dark-vortex
mode with $l=1$, $\protect\mu =1$ and $m=1$, cf. Eq. (\protect\ref%
{chi_perturbed}), produced by the numerical solution of the BdG equations,
vs. strength $U_{0}$ of the pulling-to-the-center potential (\protect\ref{U}%
). The strength of the HO trapping potential is fixed as $k=0.02$. Panels
(a) and (b) show two instability regions, which are, approximately, $%
0.01<U_{0}<0.17$ and $0.43<U_{0}<1$. }
\label{fig10}
\end{figure}

In interval (\ref{01}) corresponding to $l=1$, i.e., $0<U_{0}<1$, the
calculation of stability eigenvalues produced by the numerical solution of
the BdG equations produces a complex structure (feasibly, a fractal one) of
alternating stability and instability windows, as shown in Fig. \ref{fig10}.
This structure is confined to two sub-intervals, \textit{viz}., $%
0.01<U_{0}<0.17$ and $0.43<U_{0}<1$. 

The complex structure observed in Fig. \ref{fig10} may be interpreted as a
result of interplay and possible resonance between two basic oscillatory
modes admitted by the GP equation (\ref{basic}). This possibility is
suggested by the affinity of the dynamical setting under the consideration
to a similar 1D system, in which the counterpart of the dark vortex state
with $l=1$ is the dark soliton \cite{Dimitri}. In the 1D case, one basic
mode represents collective dipole oscillations of the condensate as a whole
(\textquotedblleft sloshing") in the HO trap \cite{dipole-mode1,dipole-mode2}%
, while the other mode amounts to oscillatory motion of the dark soliton
under the action of the same trapping potential \cite{dark-sol1,dark-sol2}.
The possibility of producing complex dynamics by the interplay of these 1D
modes was demonstrated in various setups \cite%
{sol-dipole1,sol-dipole2,Newcastle}. In the present 2D situation, perusal of
the simulations demonstrates that the quasi-spiral motion of the vortex'
pivot, in the case of the instability of the stationary vortex, is indeed
coupled to sloshing-rotating motion of the trapped condensate as a whole,
therefore the resonance between the circular motion of the vortex displaced
from the center and the collective oscillatory-rotational motion of the
condensate is a feasible cause of the complex structure displayed in Fig. %
\ref{fig10}. Detailed analysis of this conjecture should be a subject of a
separate work.

\begin{figure}[tbph]
\centering\subfigure[]{\includegraphics[width=6.5cm]{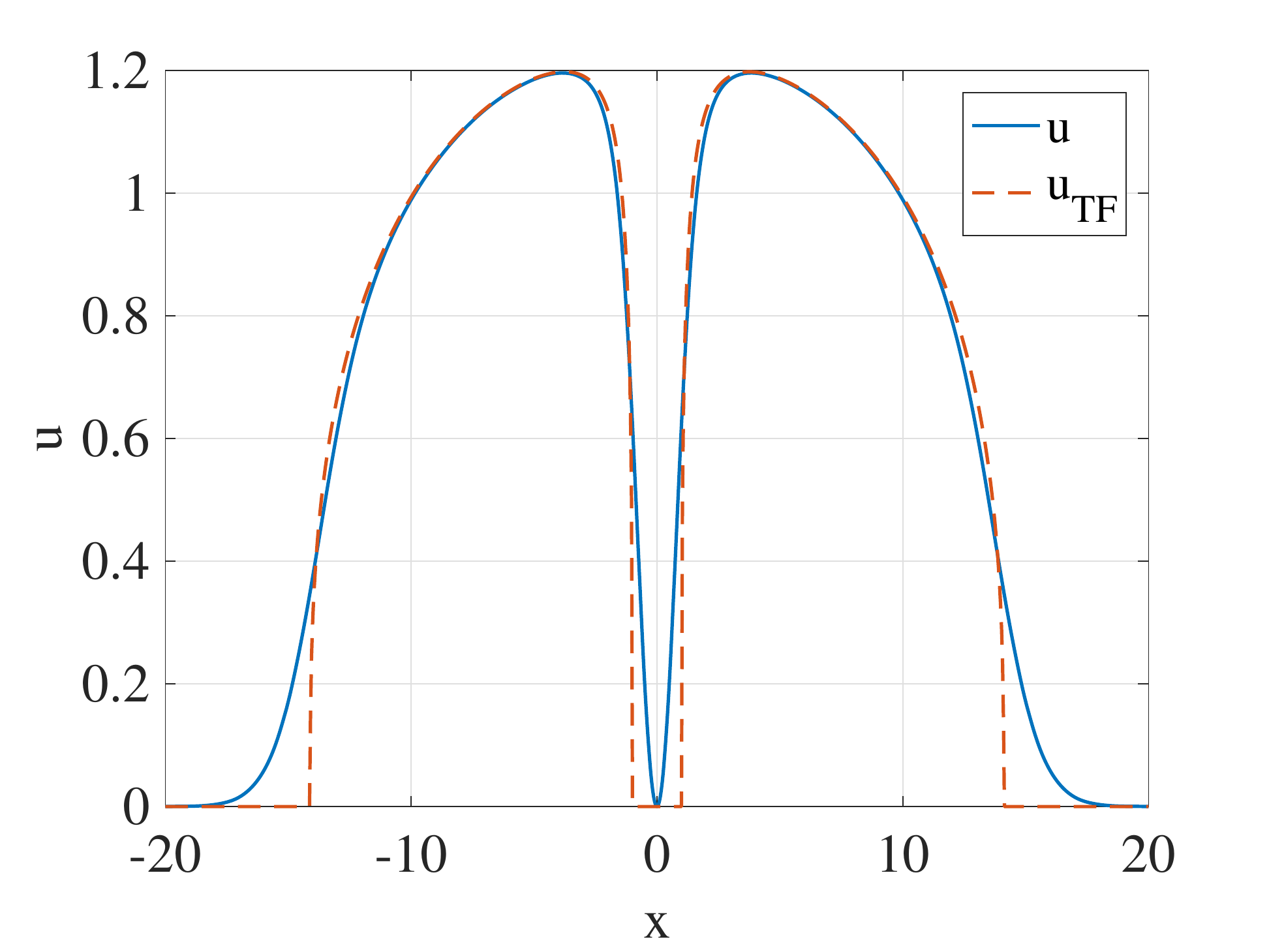}} %
\subfigure[]{\includegraphics[width=6.5cm]{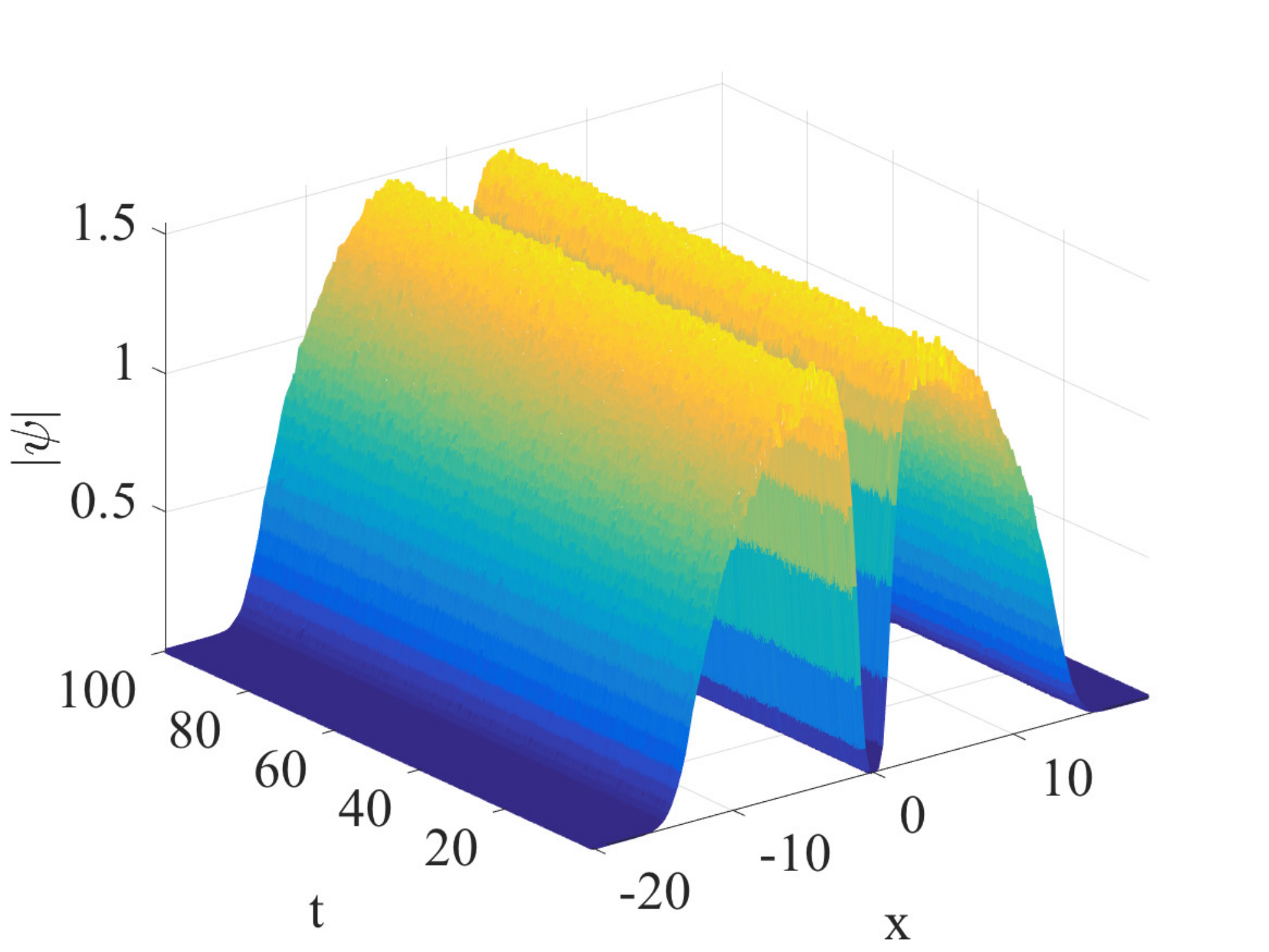}}
\caption{(a) The numerically found profile of a stable dark vortex with $l=2$
and $\protect\mu =2$, obtained at $U_{0}=0$ (hence, $U_{l}=-4$), $k=0.02$,
and $\protect\alpha =3$ in Eq. (\protect\ref{basic}). The TF counterpart of
this profile, produced by Eqs. (\protect\ref{TF2}) and (\protect\ref{rmax}),
is shown too. (b) Stable evolution of this double vortex with initially
added random perturbations.}
\label{fig11}
\end{figure}

Another essential point of the analysis is instability of double dark
vortices, with $l=2$, against splitting into unitary ones, which occurs in
many models supporting dark vortex modes \cite{HanPu, Neu, Kawaguchi,
Finland, Delgado, Poland}. In this case too, the HO trapping potential
should be kept in Eq. (\ref{basic}), to secure the robustness of the
numerical scheme (cf. a similar situation in the case of the 2D GP equation
with the cubic term \cite{HanPu}). A weak trap with $k=0.02$ is sufficient
to support stable double vortices at particular values of $U_{0}$ and $\mu $%
, provided that $U_{l}$ is negative, see Eq. (\ref{Ul}). An example of the
profile of a stable dark double vortex and its stable perturbed propagation
is presented in Fig. \ref{fig11}.

\begin{figure}[tbph]
\centering\subfigure[]{\includegraphics[width=6.5cm]{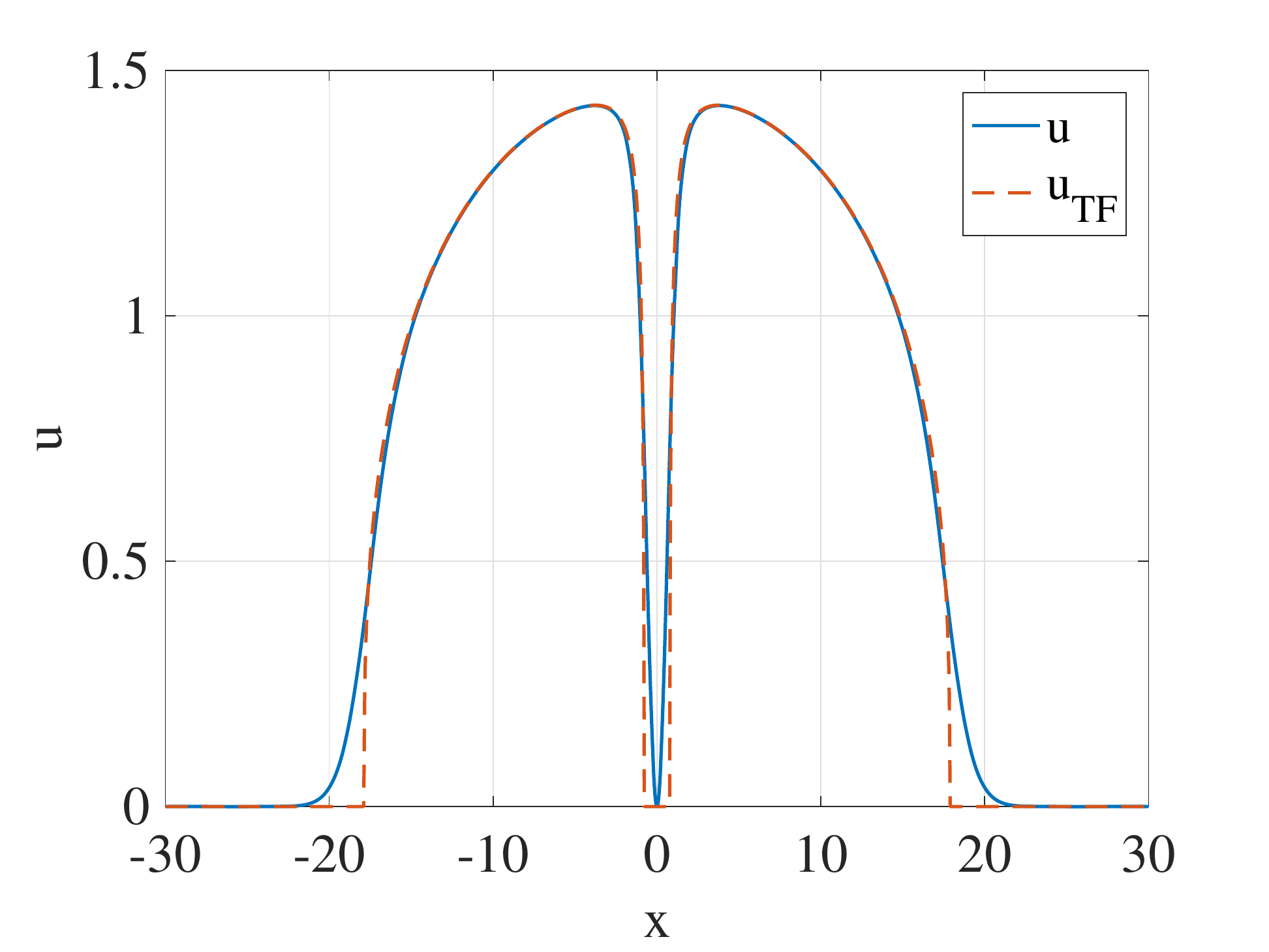}} %
\subfigure[]{\includegraphics[width=6.5cm]{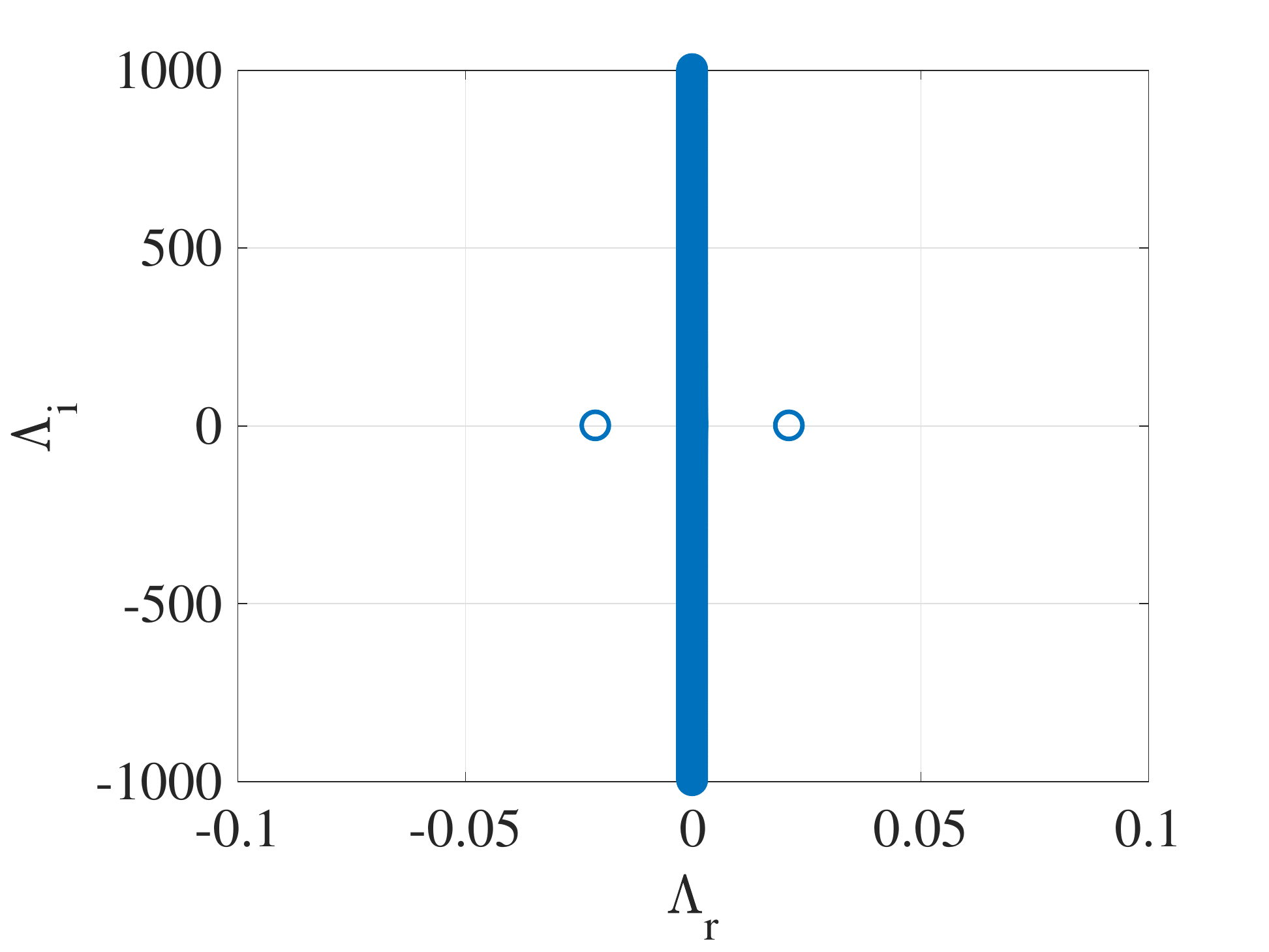}}
\caption{The same as in Figs. \protect\ref{fig7}(a,b), but for the unstable
double dark vortex ($l=2$), with $\protect\mu =3.2$. The parameters are the
same as in Fig. \protect\ref{fig11}: $U_{0}=0$, $k=0.02$, $\protect\alpha =3$%
. The unstable pair of eigenvalues corresponds to the azimuthal perturbation
index $m=2$, cf. Eq. (\protect\ref{chi_perturbed}). There are no unstable
eigenvalues with $m=1$. }
\label{fig12}
\end{figure}

The increase of the chemical potential from $\mu =2$, corresponding to the
stable double vortex in Fig. \ref{fig11}, to $\mu =3.2$ makes the dark
double vortex unstable against perturbations with azimuthal index $m=2$, cf.
Eq. (\ref{chi_perturbed}). The vortex' profile and spectrum of the
respective (in)stability eigenvalues are displayed in Fig. \ref{fig12}.

Because the instability of the double vortex in Fig. \ref{fig12} is driven
by the perturbation eigenmode with azimuthal index $m=2$, it splits the
initial vortex ring into a pair of unitary vortices. As shown in Fig. \ref%
{fig13}, the splitting is followed by recombination back into the original
vortex, thus initiating a periodic chain of splittings and recombinations.
Simultaneously, the vortex pair in the split state rotates persistently.

\begin{figure}[tbph]
\centering\includegraphics[width=14cm]{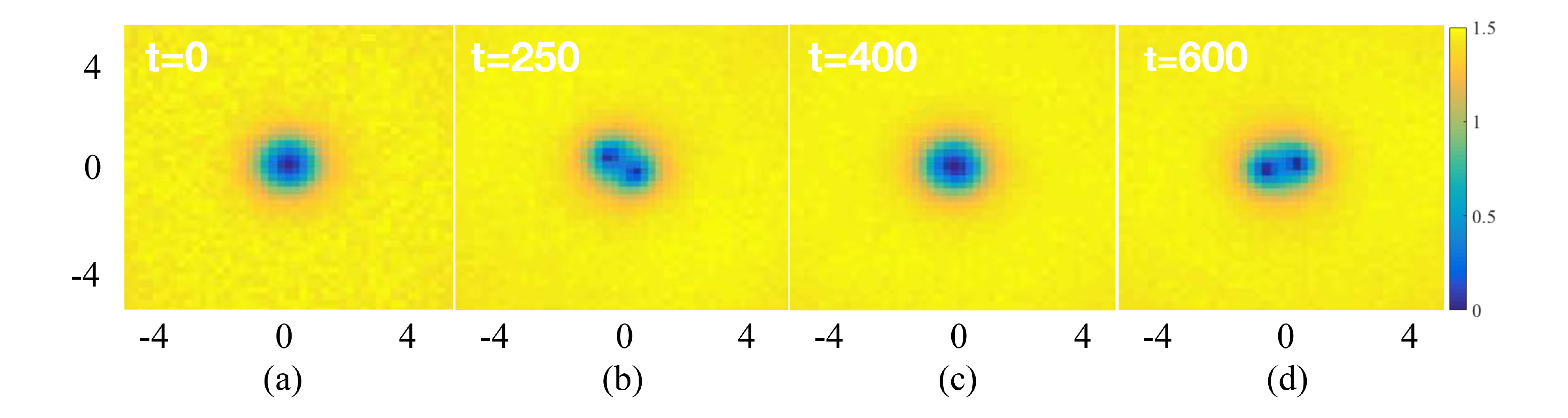}
\caption{A set of snapshots illustrating the evolution of the unstable dark
double vortex from Fig. \protect\ref{fig12}. It features a periodic chain of
splittings into a rotating pair of unitary vortices (as observed at $t=250$
and $600$), alternating with recombinations back into the double vortex (as
seen at $t=0$ and $400$). }
\label{fig13}
\end{figure}

Unlike the motion of the unstable vortex with $l=1$, which is shown above in
Figs. \ref{fig7}(c,d), the periodic fission-fusion evolution of the double
vortex, presented in Fig. \ref{fig13}, is not coupled to conspicuous
sloshing motion of the condensate as a whole. This circumstance may be
explained by the fact that monopole modes of stirring perturbations,
produced by the motion of two unitary vortices (separated by a relatively
small distance in Fig. \ref{fig13}) in opposite directions, cancel each
other through destructive interference. The remaining dipole stirring mode
produces a much weaker sloshing effect.

\begin{figure}[tbph]
\centering\subfigure[]{\includegraphics[width=8cm]{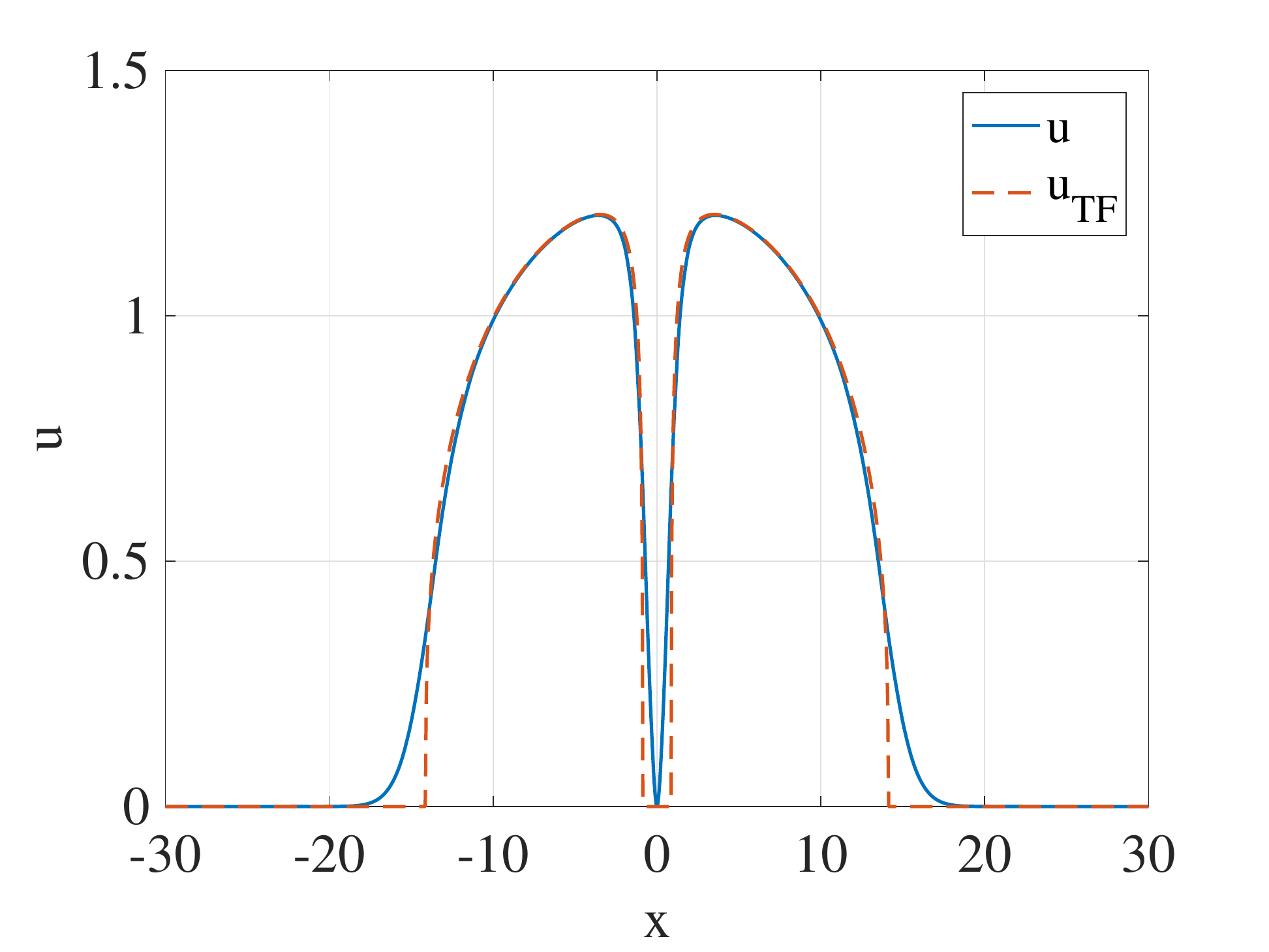}} %
\subfigure[]{\includegraphics[width=8cm]{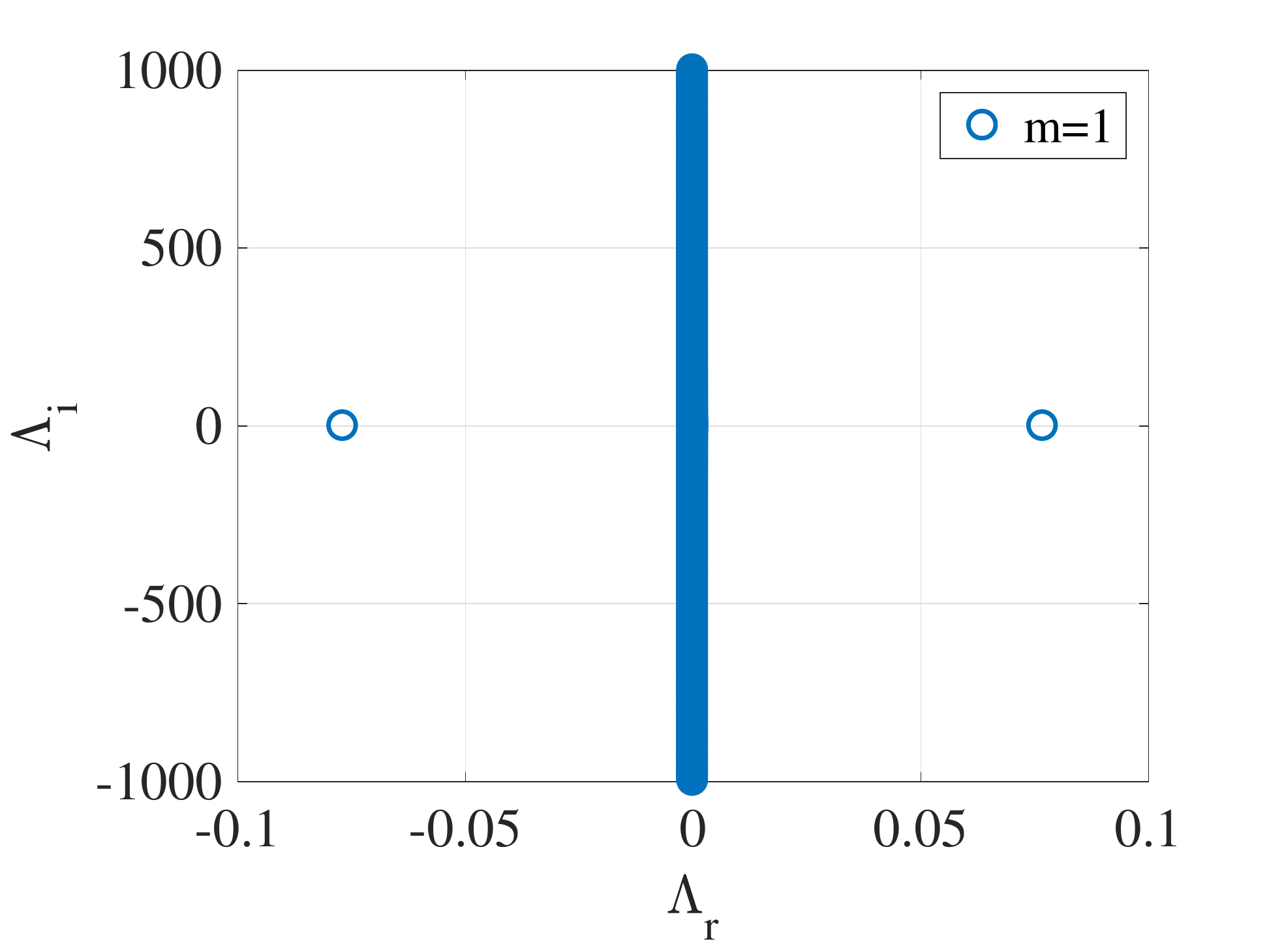}}
\caption{The same as in Fig. \protect\ref{fig12}, but for the unstable
double dark vortex ($l=2$), with $\protect\mu =2$, $U_{0}=1$, $k=0.02$, and $%
\protect\alpha =3$. (b) The spectrum of stability eigenvalues, the unstable
real pair corresponding to the azimuthal perturbation index $m=1$, cf. Eq. (%
\protect\ref{chi_perturbed}). There are no unstable eigenvalues for $m=2$. }
\label{fig14}
\end{figure}

\begin{figure}[tbph]
\centering\includegraphics[width=15cm]{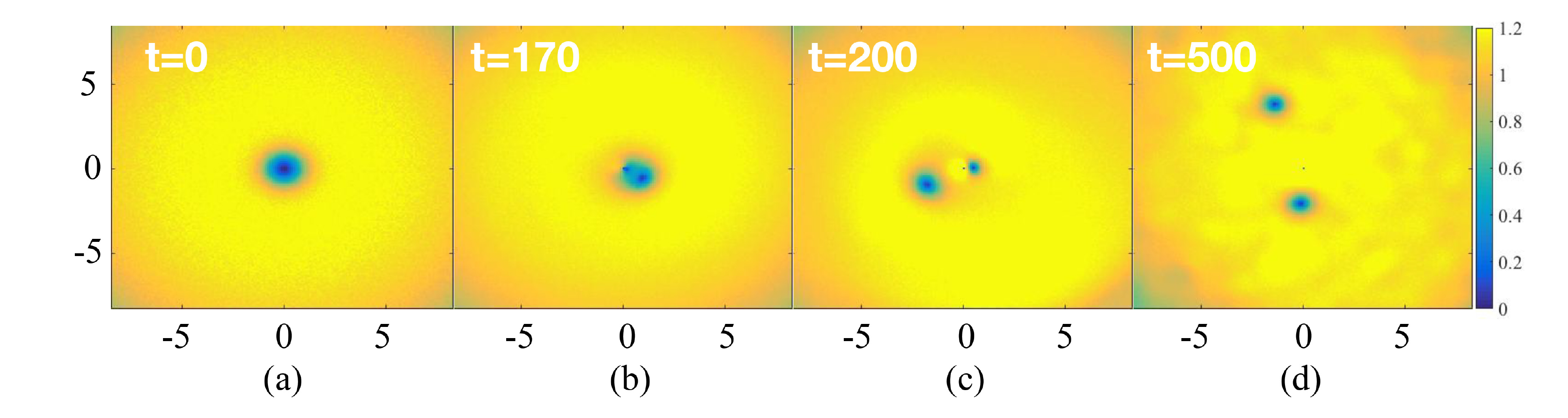}
\caption{A set of snapshots illustrating the evolution of the unstable dark
double vortex from Fig. \protect\ref{fig14}. It splits in two separate
unitary vortices, which are expelled from the center one by one, performing
circular motion in a confined region. }
\label{fig15}
\end{figure}

At other parameter values, another instability eigenmode of a double vortex
is possible. An example of such a vortex is shown in Fig. \ref{fig14}. In
this case, the instability corresponds to azimuthal index $m=1$ (rather than
$m=2$). Accordingly, direct simulations displayed in Fig. \ref{fig15} show
that the double vortex splits in a pair of unitary vortices. One of them is
expelled from the central position first, which is later followed by the
expulsion of the second one.. Both secondary vortices perform persistent
orbital motion in a confined region. Unlike what is observed in Fig. \ref%
{fig13}, the splitting is irreversible in Fig. \ref{fig15} i.e., the unitary
vortices never recombine back into a single double vortex. 


\begin{figure}[tbph]
\centering\subfigure[]{\includegraphics[width=8cm]{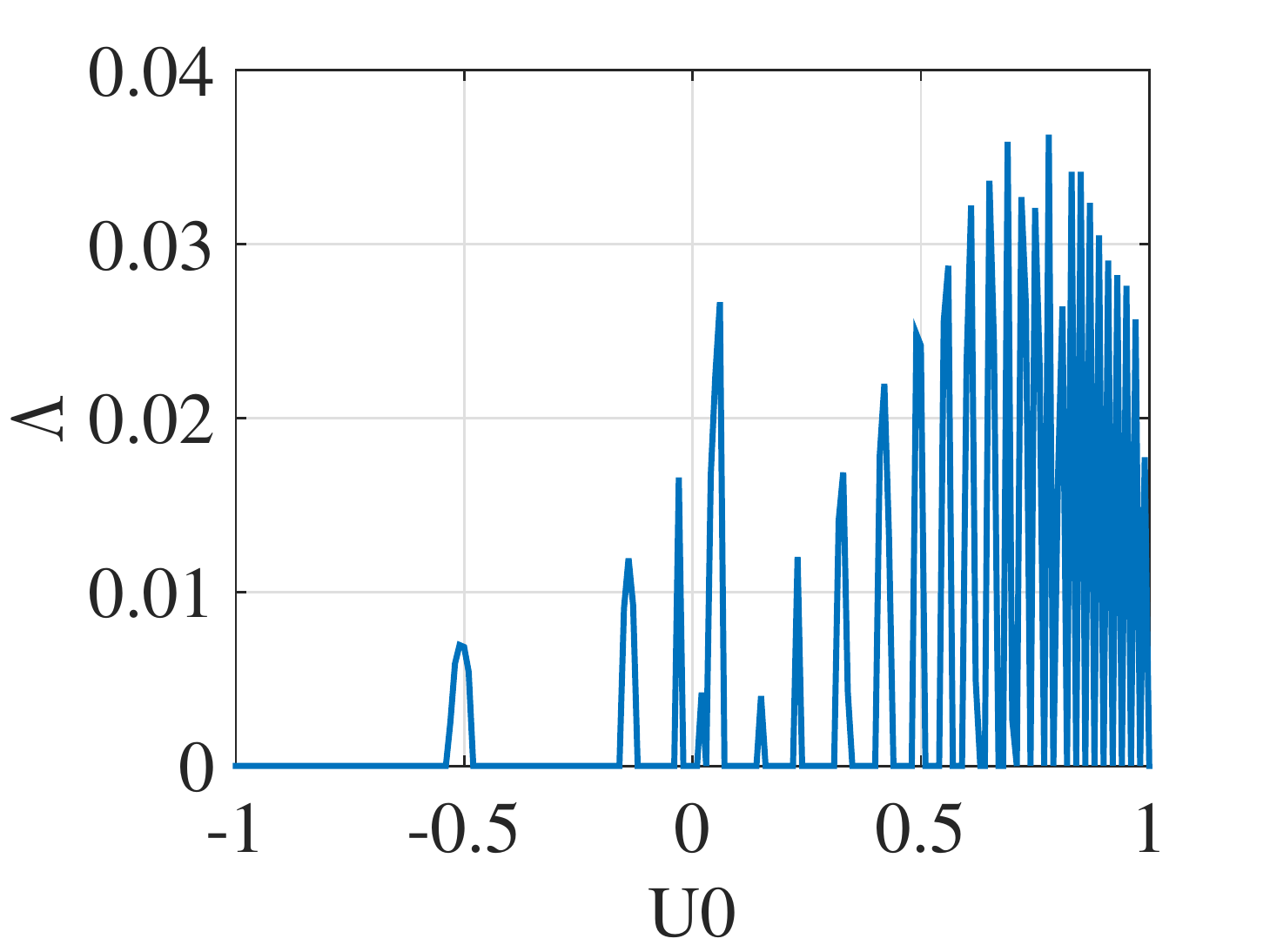}} %
\subfigure[]{\includegraphics[width=8cm]{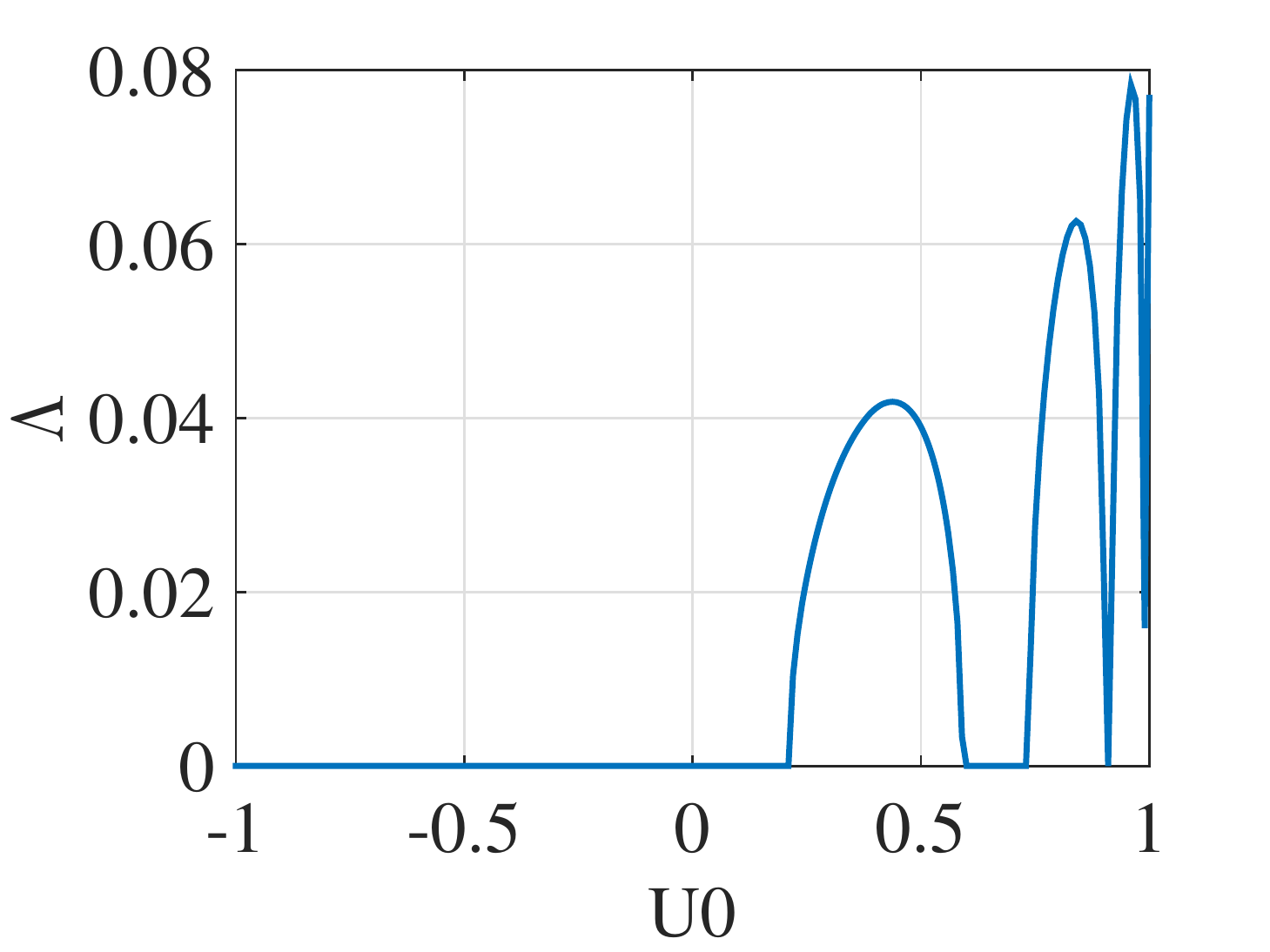}}
\caption{The instability growth rate for the dark double vortex vs. strength
$U_{0}$ of the central potential (\protect\ref{U}), at fixed values of the
trapping potential in Eq. (\protect\ref{basic}), $k=0.02$ (with $\protect%
\alpha =3$), and chemical potential, $\protect\mu =2$, for azimuthal
perturbation indices $m=2$ (a), and $m=1$ (b), cf. Eq. (\protect\ref%
{chi_perturbed}). }
\label{fig16}
\end{figure}

The alternation of stability and instability regions for the double
vortices, following the variation of $U_{0}$, forms a complex structure,
which is plotted in Fig. \ref{fig16}, separately for eigenmodes with the
azimuthal index $m=2$ in (a), and $m=1$ in (b). Note that there is no
instability at $U_{0}<-1/2$, and the dark double vortex does not exist at $%
U_{0}\geq 4$. The structure observed in the figure may be interpreted as a
result of the interplay and possible resonance between two oscillatory
modes. One represents internal oscillations in the pair of separated unitary
vortices, while the other mode corresponds to rotation of the vortex pair.

\section{Conclusion}

The main objective of this work is to extend the family of singular but
physically relevant singular vortices in the 2D models of optical and matter
waves, which combine attractive potential (\ref{U}) and the quartic ($\alpha
=3$) or quintic ($\alpha =4$) self-repulsion in Eq. (\ref{basic}), for the
antidark vortex states built on top of a finite background. It is
demonstrated that the TF (Thomas-Fermi) approximation provides a very
accurate fit for the numerically found singular states. Their stability
exactly follows the recently proposed analytical condition, given by Eq. (%
\ref{as_before}) for $\alpha =3$, and by the newly derived Eq. (\ref{gamma4}%
) for $\alpha =4$. A nontrivial finding is the existence of the singular
states in the case when the effective potential strength, given by Eq. (\ref%
{Ul}), is negative, corresponding to weak repulsion, instead of the pull to
the\ center. In this case, the shape of the mode features a shallow local
minimum, as shown in Fig. \ref{fig3}(c). An essential novel result is the
scenario for the excitation of perturbed singular vortices by the ordinary
(nonsingular) vortex input, which is shown in Fig. \ref{excitation}.
Parallel to that, the analysis is performed for the usual regular (dark)
vortices, which are also supported by a finite background, in the case when
the effective central potential is repulsive. The dark states with
vorticities $l=0$ and $1~$are completely stable for $U_{0}<0$ in Eq. (\ref{U}%
)). In the interval of $0<U_{0}<1$, where the effective potential's
strength, including the centrifugal term, $U_{0}-1$, corresponds to the
repulsion, the accurate stability analysis for the dark vortex with $l=1$ is
possible in the presence of the weak confining HO (harmonic-oscillator)
potential. In this case, the vortex features an intricate pattern of
alternating stability and instability windows, as shown in Fig. \ref{fig10}.
Unstable vortices spontaneously move along complex trajectories, which do
not leave the central area confined by the HO trap (an example is displayed
in Fig. \ref{fig7} (c) and (d)). The presence of the weak trap is also
necessary for the accurate analysis of the stability of dark double
vortices. In this case too, stability and instability regions compose a
complex pattern. Unstable double vortices periodically split into rotating
pairs of unitary vortices and recombine back. Finally, simple but new
solutions for flat vortices are obtained at the border between the antidark
singular vortices and dark regular ones.

A challenging direction for extension of the present work may be the
analysis of antidark vortex states in the 3D model with the same central
potential.

\section*{Acknowledgment}

This work was partially supported by grant No. 1286/17 provided by the
Israel Science Foundation. Z.C. acknowledges an excellence scholarship
provided by the Tel Aviv University.

\end{document}